\documentclass[useAMS,usenatbib]{mnras}
\usepackage{graphicx}
\usepackage{amsmath}
\usepackage{amssymb}
\usepackage{comment}
\usepackage{subfigure}
\usepackage[normalem]{ulem}

\include{defns}
\def\gsim{\;\rlap{\lower 2.5pt
 \hbox{$\sim$}}\raise 1.5pt\hbox{$>$}\;}
\def\lsim{\;\rlap{\lower 2.5pt
   \hbox{$\sim$}}\raise 1.5pt\hbox{$<$}\;}
\defcitealias{liu09a}{Liu et al. (2009a)}
\defcitealias{liu09b}{Liu et al (2009b)}

\newcommand{\tr}[1]{\textrm{#1}}
\newcommand{\ee}[1]{\times10^{#1}}
\newcommand{\dd}[2]{\frac{\textrm{d}#1}{\textrm{d}#2}}
\newcommand{\pp}[2]{\frac{\partial#1}{\partial#2}}
\newcommand{\Lscript}{\mathcal L}

\title[]{Interaction of Cosmic Rays with Cold Clouds in Galactic Halos}
\author[Wiener, Oh, \& Zweibel]{Joshua Wiener$^{1,2}$, S. Peng Oh$^{1}$, \& Ellen G. Zweibel$^{2,3}$\\
$^{1}$ Department of Physics; University of California; Santa Barbara, CA 93106, USA.\\
$^{2}$ Department of Astronomy, University of Wisconsin-Madison, Madison, WI 53706, USA.\\
$^{3}$ Department of Physics, University of Wisconsin-Madison, Madison, WI 53706, USA.}

\begin{document}
\bibliographystyle{mnras}

\pagerange{000--000} \pubyear{0000}
\maketitle

\label{firstpage}

\begin{abstract}
We investigate the effects of cosmic ray (CR) dynamics on cold, dense clouds embedded in a hot, tenuous galactic halo. If the magnetic field does not increase too much inside the cloud, the local reduction in Alfv\'en speed imposes a bottleneck on CRs streaming out from the star-forming galactic disk. The bottleneck flattens the upstream CR gradient in the hot gas, implying that multi-phase structure could have global effects on CR driven winds. A large CR pressure gradient can also develop on the outward-facing edge of the cloud. This pressure gradient has two independent effects. The CRs push the cloud upward, imparting it with momentum. On smaller scales, the CRs pressurize cold gas in the fronts, reducing its density, consistent with the low densities of cold gas inferred in recent COS observations of local $L_{*}$ galaxies. They also heat the material at the cloud edge, broadening the cloud-halo interface and causing an observable change in interface ionic abundances. Due to the much weaker temperature dependence of cosmic ray heating relative to thermal conductive heating, CR mediated fronts have a higher ratio of low to high ions compared to conduction fronts, in better agreement with observations. We investigate these effects separately using 1D simulations and analytic techniques.
\end{abstract}

\section{Introduction}

Cosmic rays (CRs) in our Galaxy are remarkably isotropic (to several parts in $\sim 10^{4}$), despite the discreteness and transience of supernova sources. Analysis of CR spallation products and radio nuclide abundances indicate residence lifetimes of $\sim 10^{7}$ yr (at 1 GeV), orders of magnitude larger than the light travel time across the Galaxy. These facts led to the `self-confinement' theory for CRs, in which CRs streaming super-Alfv\'enically excite a gyro-resonance instability, amplifying small-scale magnetic fluctuations which efficiently scatter the CRs \citep{wentzel69,kulsrud69,kulsrud05,zweibel13}. The CRs are rendered nearly isotropic in the Alfv\'en wave frame and thus advect with velocity $u + v_A$ (where $u,v_A \ll c$ are the gas and Alfv\'en velocity respectively). In addition, CRs diffuse relative to the wave frame; if wave damping is strong, diffusion coefficients are larger. For most of the ISM, the assumption that diffusion is negligible and thus that CRs stream at the Alfv\'en speed relative to the gas is an excellent approximation \citep{skilling71}. The wave-particle interactions result in a short ($\sim$pc) mean free path, which means that CRs can be treated as a fluid. Wave-particle interactions also allow CRs to transfer momentum (with force $\nabla P_c$) and energy (at a rate $v_A \cdot \nabla P_c$) to the gas. In the interstellar medium (ISM), where the CRs are comparable in energy density to the thermal gas, these rates can be considerable.

It was noted early on that CRs can drive galactic winds \citep{ipavich75,breitschwerdt91}, both directly, and by heating the thermal gas. A wind in the inner Milky Way driven by a combination of thermal and CR pressure is a good match to synchrotron and soft X-ray measurements \citep{everett08,everett10}. Recently, there has been a burst of activity on CR-driven winds as a feedback mechanism, both analytically \citep{socrates08}, and numerically with isotropic \citep{uhlig12,booth13,salem14,simpson16,wiener16} and field-aligned \citep{hanasz13,girichidis16,ruszkowski16,pakmor16} CR transport.

CR driven winds have a number of attractive properties. With no fine-tuning, the simulations obtain the mass-loading factors required to reproduce the faint end of the luminosity function, and reproduce the observed trend of wind velocity with galaxy circular velocity. Unlike supernova-driven outflows, CRs do not lose energy catastrophically to radiative cooling in dense regions, but gently push a massive, multi-phase wind on scales comparable to galactic scale heights. In good agreement with observations (and in contrast to models of thermally driven winds), CR winds have velocities that rise with distance (cf \citet{steidel10}), and entrain substantial warm ionized, $\sim$10$^{4}$K gas \citep{chen10,rubin10,tumlinson11}. Indeed, CR-driven winds are significantly cooler than thermal winds (although efficient cosmic ray-thermal gas coupling generally requires that the gas be almost fully ionized). While momentum transfer from the CRs is generally thought to be the important factor in driving winds, the effects of CRs on thermal pressure gradients (through CR heating) can be comparable (e.g. \citealt{everett08}).  

Heating by CRs in the ISM is usually confined to considerations of Coulomb and hadronic heating, even though the potential importance of CR wave heating was recognized early on \citep{skilling71, wentzel74}. In the ICM, wave heating from CRs injected by a central AGN has been proposed as a means of stemming cooling flows \citep{loewenstein91,guo08-CR,fujita11}. CR heating balancing cooling is consistent with the smooth gamma-ray and radio emission in M87 \citep{pfrommer13}, which suggests that the CRs are well mixed. In the ISM, CR wave heating potentially provides the required supplemental heating to explain observed line ratios \citep{reynolds99} in the warm ionized medium of the Galaxy \citep{wiener13}. 

An important piece missing from all these calculations is the effect of multi-phase structure. The characteristic e-folding scale on which CRs transfer momentum and energy is the CR scale height; its large ($\sim$ kpc) scale is what enables the gentle, distributed forces and heat which make CR feedback so effective. However, multi-phase structure causes sharp gradients in Alfv\'en speed due to sharp density contrasts. In the limit of effective wave-particle scattering where the CRs are tightly locked to the gas, it can be shown that $P_c \propto v_A^{-\gamma_c}$ \citep{breitschwerdt91}, or $P_c \propto \rho^{\gamma_c/2}$, since observationally the B-field does not appear to show significant density dependence below $n \lsim 300 \, {\rm cm^{-3}}$ \citep{crutcher10}. While ion-neutral damping of MHD waves in dense neutral clouds can decouple the CRs and gas \citep{kulsrud69,everett11}, and allow free-streaming within the cloud, coupling will always be re-established when CRs exit the cloud back into fully ionized gas. {\it In this case, the CR scale height is of order the (very short) density scale height of the interface, and it transfers energy and momentum on this short length scale.} This implies that multi-phase structure has important effects on the spatial footprint of CRs.  

Warm clouds can also have a more subtle effect, discussed qualitatively by \citet{skilling71} and \citet{begelman95}, but otherwise never explored or calculated in detail. It exploits two important features of CR streaming: CRs can only stream down their density gradient (within the framework of the fluid model), and the streaming instability which couples the CRs to the gas is only triggered when the bulk drift speed $v_{\rm D} > v_A$. Since $v_{\rm D} \sim v_A \propto \rho^{-1/2}$, a cloud of warm ($T\sim 10^{4}$K) ionized gas embedded in hot ($T\sim 10^{6}$K) gas results in a minimum in drift speed. This produces a `bottleneck' for the CRs; CR density is enhanced as they are forced to slow down, akin to a traffic jam. Since CRs cannot stream up a gradient, the system readjusts to a state where the CR profile is flat up to the minimum in $v_A$; thereafter the CR pressure falls again {\bf as $v_A^{-\gamma_c}$}. Importantly, since the CRs in the hot gas preceding the cloud now have a bulk drift speed less than the local critical drift speed for triggering the streaming instability, {\it they are no longer coupled to the gas}, and can no longer exert pressure forces or heat the gas. Instead, they do so only on the back side of the cloud, exerting pressure forces and heating which have important effects on cloud structure. Thus, the presence of warm gas upstream in the halo can influence CR driving and heating in the disk. 

This paper is an exploratory study of these effects. It examines the dynamic effects of CRs on cold clouds on two separate length scales - the large scale pushing of the cloud by the CR pressure gradient, and the much smaller scale thickening of the cold/hot gas interface by CR wave heating. We use simple 1D simulations to study the first effect, and derive simple 1D steady-state models to study the second. The outline of the paper is as follows. In \S \ref{sec:equations} we introduce the evolution equations for CRs and for the thermal gas that will be relevant in our models. \S \ref{sec:ZEUS} describes the time-dependent simulations that demonstrate the formation of the bottleneck (to our knowledge, the first time this long predicted phenomenon is shown to be the outcome of an evolutionary process) and simulate the pushing action, while \S \ref{sec:steadystate} explains the steady-state models for the cloud interface thickened by CR heating. The validity of the fluid equations used in our model is justified in \S\ref{sec:frame_lock}. We conclude in \S\ref{sec:conclusions}.

\section{Evolution Equations}\label{sec:equations}
\subsection{CR Dynamics}\label{sec:CRequations}
In the fluid approximation, the bulk properties of the CR population are governed by the CR transport equation {\citep{mckenzie82,breitschwerdt91,guo08-CR}: 
\begin{equation}\label{CR1}
\pp{P_c}{t}=(\gamma_c-1)(\mathbf{u}+\mathbf{v_A})\cdot\nabla P_c -\nabla\cdot \mathbf{F_c}+Q,
\end{equation}
\[
\mathbf{F_c}=\gamma_c P_c(\mathbf{u}+\mathbf{v_A})-\mathbf{b}\kappa_c(\mathbf{b}\cdot\nabla P_c).
\]
Here, $P_c$ is the CR pressure, $\gamma_c=4/3$ is the adiabatic index of the CRs (assumed to be relativistic), $\mathbf{u}$ and $\mathbf{v_A}$ are the local gas and Alfv\'en velocities respectively, and $Q$ contains any CR sources and sinks. $\mathbf{F_c}$ represents the CR energy flux, and the CR diffusion coefficient $\kappa_c$ is the anisotropic diffusion coefficient of the CRs relative to the wave frame, along the local magnetic field direction $\mathbf{b}$. 

Consider a simple case where we have no sources or sinks, and no CR diffusion. That is, the CRs are locked to the wave frame everywhere. Let's also assume the bulk gas motion is negligible, so $u=0$. Then \eqref{CR1} reduces to
\begin{equation}
\pp{P_c}{t}=-\mathbf{v_A}\cdot\nabla P_c - \gamma_cP_c\nabla\cdot \mathbf{v_A}.
\end{equation}
If we further consider a one-dimensional, plane-symmetric system, this becomes
\begin{equation}
\pp{P_c}{t}=-v_A\dd{P_c}{z}-\gamma_cP_c\dd{v_A}{z}.
\end{equation}
This has a steady-state solution:
\[
0=-v_A\dd{P_c}{z}-\gamma_cP_c\dd{v_A}{z}
\]
\[
\frac{\tr{d}P_c}{P_c}=-\gamma_c\frac{\tr{d}v_A}{v_A}
\]
\begin{equation}\label{CR2}
P_cv_A^{\gamma_c}=\tr{constant}.
\end{equation}

We caution that this solution assumes that waves are present regardless of the spatial variation of $P_c$. If there is some external source of waves that is driving the CR population in one coherent direction through the plasma, then \eqref{CR2} is valid as written. Note that one can have CRs streaming up their own gradient in this case, but this is physically reasonable since the CRs are being driven by the external wave source\footnote{Note that while CRs can scatter off compressible fast modes \citep{yan02}, external Alfv\'enic turbulence is highly anisotropic at small scales and scatters CRs inefficiently \citep{chandran00}.}.

If no such source exists, and only CR gradients produce Alfv\'en waves via the streaming instability, then we must take care when using \eqref{CR2}. This is because the direction of $v_A$ must now always point down the CR gradient. As such $P_c$ and $v_A$ are not entirely independent. Note that \eqref{CR2} is perfectly fine so long as the solution is monotonic. Problems arise when there is a minimum in $|v_A|=B/\sqrt{4\pi\rho_g}$. Then the solution to \eqref{CR2} may have $v_A$ pointing \emph{up} the CR gradient, which is a contradiction.

What happens in this case? Suppose the bulk flow of CRs is to the right\footnote{We distinguish here between the bulk flow velocity, which is the velocity averaged over all cosmic rays, and the speed of individual particles, which is close to $c$.}, at a speed $|v_A|$ which varies in space. At some critical point in this flow there is a local minimum in $|v_A|$, say $v_{A,\tr{min}}$. As the CRs approach this minimum, their flow speed is decreasing, and so $P_c$ is increasing as we move to the right. But this means the gradient of $P_c$ now points to the left -- the CR flow changes direction. Now there is an increasing flow speed to the left, so $P_c$ at the critical point begins to fall. The equilibrium state is one where the upstream CRs move with the same bulk speed $v_{A,\tr{min}}$, and have a constant, spatially uniform energy density $P_c$. Upstream of the bottleneck, the CR bulk speed is less than the local Alfv\'en speed, and so the streaming instability is not activated and no Alfv\'en waves are present in this region. Downstream of the critical point, where $|v_A|$ is increasing towards the right, there is a gradient in $P_c$, Alfv\'en waves are generated, and \eqref{CR2} holds. The situation is analogous to a traffic jam -- drivers are stuck moving slower than they normally would because there are other cars ahead of it that are in the way. Figures \ref{case1} and \ref{case2a} show, respectively, a cosmic ray front advancing and reaching an equilibrium in which $P_cv_A^{\gamma_c}$ is constant, and the formation of a cosmic ray pressure plateau and re-engagement induced by a thermal gas density maximum.

Why exactly does \eqref{CR2} fail? The reason is that \eqref{CR2} assumes there are Alfv\'en waves present everywhere. For any monotonic solution this assumption holds. But for any solution where the gradient of $P_c$ vanishes anywhere, this assumption is contradicted. The true solution is as described above. In the region upstream of the critical point, where $P_c$ is flat, there simply are no Alfv\'en waves since there is no CR gradient to drive them. Note that although the CRs are not flowing at the local value of $|v_A|=B/\sqrt{4\pi\rho_g}$, there is no diffusion. Diffusion describes drift of the CR population relative to the local wave frame. In the absence of any waves, ``diffusion'' has no real meaning.

\subsection{Gas Dynamics}
The typical evolution equations for a thermal gas are determined from mass, momentum, and energy conservation. In one dimension they are:
\begin{equation}\label{eq:gas1}
\pp{\rho}{t}+\pp{(\rho u)}{z}=0
\end{equation}
\begin{equation}\label{eq:gas2}
\pp{(\rho u)}{t}+\pp{}{z}(\rho u^2+P_g+P_c)=0
\end{equation}
\begin{equation}\label{eq:gas3}
\pp{E_g}{t}+\pp{}{z}\left[(E_g+P_g)u-\kappa_g\pp{T}{z}\right]=-\rho\mathcal{L}-v_A\pp{P_c}{z}
\end{equation}
In the above, $\rho$ and $u$ are the gas density and velocity, $P_g$ and $E_g$ are the gas thermal pressure and gas \emph{total} energy density $E_g=P_g/(\gamma_g-1)+1/2\rho v^2$, and $\kappa_g$ is the thermal conduction coefficient. $P_c$ represents the CR pressure, and $\rho\mathcal{L}$ represents the net cooling function. The effect of CRs is contained in a pressure gradient term in the momentum equation and a heating term in the energy equation. The heating rate due to CR pressure gradients is derived in \cite{wiener13a}.

Our time-dependent simulations use various simplifications of the above set of equations. Our simplest test cases only evolve the CRs, while gas properties are held fixed. For computational reasons we ignore thermal conduction and radiative losses in our other simulations, $\kappa_g = 0$, $\rho\mathcal{L} = 0$ (see \S \ref{sec:regimes} for further details).

\section{Hydrodynamic Simulation}\label{sec:ZEUS}
We use a 1D version of the MHD simulation code {\small ZEUS3D}, adapted to include cosmic ray dynamics. Equation \eqref{CR1} is integrated in time using a finite-difference method on a fixed spatial grid. In this simulation we do not keep track of any momentum-dependent quantities -- only $P_c$ is evolved.

To account for the bottleneck effect discussed in \S\ref{sec:CRequations}, we must account for direction when calculating Alfv\'en velocities. The most straightforward way to do this is to define the wave velocity in each grid cell to be
\begin{equation}
v_A=-\frac{B}{\sqrt{4\pi\rho_g}}\tr{sgn}\left(\dd{P_c}{z}\right).
\end{equation}
But this lends itself to numerical instability -- spatial oscillations will develop as CRs slosh back and forth between adjacent grid cells. This can be understood from overshoot as CRs stream in (out) of local minima (maxima). We use a graceful solution introduced by \cite{sharma10-stream}, where we replace the sign function with a smooth hyberbolic tangent function. That is, we define the wave velocity as
\begin{equation}\label{smoothing}
v_A=-\frac{B}{\sqrt{4\pi\rho_g}}\tr{Tanh}\left(\frac{\tr{d}P_c/\tr{d}z}{\epsilon}\right).
\end{equation}
for some scale value $\epsilon$. The simulation is then stable to numerical oscillations provided we use the accompanying time-step restriction
\begin{equation}
\Delta t\le \Delta x^2\epsilon/2P_c|v|
\end{equation}
at every local extremum of $P_c$. For $|v|$ we use a fixed value of $4.0\ee{7}$ cm/s, which we expect to be greater than $|v_A|$ everywhere in the simulated region at any time. For $\epsilon$ we use different values for different tests, ranging from $\epsilon=1\ee{-37}$ to $\epsilon=1\ee{-36}$ erg/cm$^4$. For the typical CR pressures of $\sim 10^{-13}$ erg cm$^{-3}$ used in these tests, this corresponds to a length scale over which the CR pressure changes of $L_\tr{CR}\sim 100$ kpc. That is, for gradients with length scales less than 100 kpc, the tanh function should closely resemble the sign function. (See \cite{wiener13} for a discussion of convergence with respect to this parameter - in that work we show our results are converged for $\epsilon$ corresponding to $L_\tr{CR}=3$ Mpc, but our simulated domains were much larger ($\sim 1$ Mpc rather than $\sim 1$ kpc in this work)).

\subsection{Test Case 1: Wave Locking}
We first test to see if the code will reach an equilibrium in the presence of a localized CR source. To do this we set up a sloped gas density profile:
\begin{equation}
n_e=n_0-\Delta n\frac{z}{1\tr{ kpc}},
\end{equation}
with $n_0=5.672\ee{-3}$ cm$^{-3}$ (this gives an Alfv\'en speed of $v_A=87$ km/s for our choice of mean molecular weight $\mu_e=1.18$ and magnetic field) and $\Delta n=9\ee{-3}$ cm$^{-3}$. Initially there is only a small uniform background of CRs. At the start of the simulation a steady CR source at the left boundary is turned on. Reflecting boundary conditions are used at the location of the CR source, and outflow conditions are used at the other boundary. We set a magnetic field of 3.26 $\mu$G throughout\footnote{We had intended to use a field of 3 $\mu$G, such that the Alfv\'en speed would be an even 80 km/s, but due to a coding error in our calculation of the Alfv\'en speed we effectively use a field of 3.26 $\mu$G.}. There is no gravitational field and no wave damping. The gas temperature is set such that the initial configuration is in pressure equilibrium. The properties of the gas are held fixed -- CR pressure feedback and wave heating are turned off. The only purpose of this simulation is to ensure that we obtain an equilibrium in a region of strictly increasing $|v_A|$. This simulation very quickly reaches an equilibrium CR profile (see figure \ref{case1}, top plot).

We can also see that in this case, which has no local minima in $|v_A|$ and hence no bottleneck effect, the solution \eqref{CR2} is correct. The bottom plot of figure \ref{case1} shows how, away from the CR source, $P_cv_A^{\gamma_c}$ is constant to within 1\% once equilibrium is reached.

\begin{figure}
\includegraphics[width=8cm,trim=1cm 2cm 2cm 0cm]{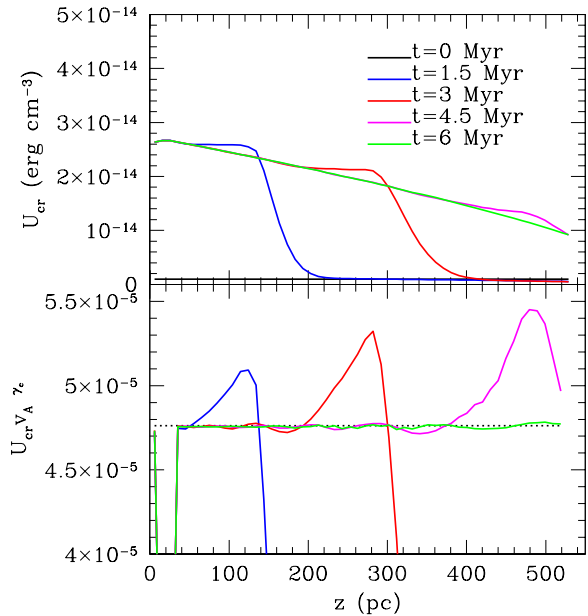}
\caption{Test Case 1: Wave Locking. A CR source is placed at the far left. Top: CR energy density $U_\tr{cr}$. Bottom: The quantity $U_{cr}v_A^{\gamma_c}$. In equilibrium, this quantity is nearly constant, as expected from the solution \eqref{CR2}. The expected value is shown as a dashed line.}\label{case1}
\end{figure}

\begin{figure}
\includegraphics[width=7cm,trim=1cm 0cm 4cm 0cm]{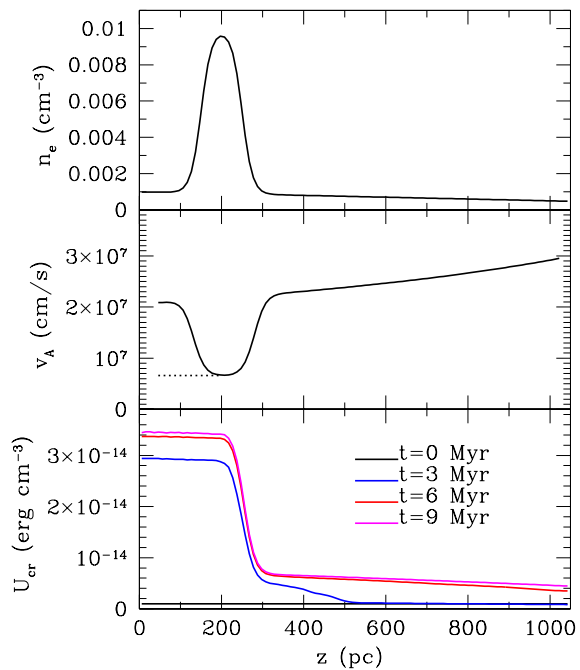}
\caption{Test Case 2: Bottleneck Effect. A simple test case. A CR source is placed at the far left, with a fixed warm cloud a short distance away. Top panel: gas density. Middle panel: Alfv\'en speed (solid) and theoretical steady-state CR bulk drift speed (dotted). Bottom: CR energy density at differernt times.}\label{case2a}
\end{figure}

\subsection{Test Case 2: Bottleneck Effect}
Let us now set up a system with a warm cloud, but not yet allow the gas to evolve. This will test if the code handles the bottleneck effect correctly. As before we use reflecting boundary conditions at the location of the CR source and outflow conditions at the other end. The density of the cloud edge is described by a Tanh function, as in \cite{everett11},
\begin{multline}
n_e=n_\tr{hot}+(n_\tr{cold}-n_\tr{hot})\times \\ 
\left(\frac{1}{2}+\frac{1}{2}\tanh\left(\frac{z-z_\tr{front}}{\Delta z}\right)\right) \times \\ 
\left(\frac{1}{2}+\frac{1}{2}\tanh\left(\frac{z_\tr{back}-z}{\Delta z}\right)\right) \\ 
-n_\tr{hot}\frac{z}{L}.
\end{multline}
with a smoothing scale of $\Delta z=$ 25 pc. The scale is chosen to be large to avoid a need for high spatial resolution for this test. The hot medium outside the cloud has temperature $10^6$ K and density $n_\tr{hot}=10^{-3}$ cm$^{-3}$. The inside of the cloud is at $10^5$ K with $n_\tr{cold}=10^{-2}$ cm$^{-3}$ (this example is purely illustrative; in what follows we adopt a more realistic temperature $T \sim 10^{4}$K for the cold gas). We again have a 3.26 $\mu$G field throughout.

Our choice of a constant magnetic field in a region with highly varying density is largely for simplification, but we can justify it if we assume the cloud in question formed by compression along the local field. For a system in rough equipartition between thermal and magnetic energies this is a reasonable assumption. We also reference observations that the B-field does not significantly vary with density in the diffuse ISM (\cite{crutcher10}, although the densities where the field is measured are higher than assumed here).

There is also a slight gradient in the gas density (with corresponding gradient in temperature to ensure equal gas pressure everywhere) characterized by length scale $L=2$ kpc. This is so that the Alfv\'en speed is not spatially uniform at the boundary. We find that when the Alfv\'en speed is uniform, the CR profile becomes flat, and when the CR profile is flat at the boundary, the smoothing scheme \eqref{smoothing} prevents CRs from leaving the domain.

Figure \ref{case2a} shows the background density, Alfv\'en speed, and the resulting CR profile. The Alfv\'en speed is a minimum at the center of the cloud, so there is a bottleneck there. The CR profile quickly reaches an equilibrium where the CR pressure is uniform upstream of the bottleneck, and follows the monotonic solution \eqref{CR1} beyond. The theoretical bulk drift speed upstream of the bottleneck is shown in the dotted line in the middle panel - in this region the bulk flow of the CRs is less than the local Alfv\'en speed, so the streaming instability is not activated and there are no waves present. To our knowledge this is the first verification of the bottleneck effect in a controlled numerical experiment.

\subsection{Test Case 3: CR pressure and heating}\label{c3}

We now include the dynamical and thermal influence of CRs, both of which are potentially important. The timescale on which momentum doubles is $t_{\rm dyn} \sim \rho u/\nabla P_c$, while the CR heating time is $t_{\rm heat} \sim U_{\rm t}/v_A \nabla P_c$. The ratio of these timescales is therefore: 
\begin{equation}
\frac{t_{\rm dyn}}{t_{\rm heat}} \sim \frac{\rho u v_A} {U_{\rm t}} \sim {\mathcal M} \beta^{-1/2} 
\end{equation}
where we have used $U_{\rm therm} \sim \rho c_{s}^{2}$. This is of order unity in the $\beta \sim 1$, transsonic ISM, and either could dominate depending on the situation.

For this simple test we use the same cloud setup as in test 2, but with a ``cool'' cloud temperature and density of $10^4$ K and $10^{-1}$ cm$^{-3}$. Additionally, we turn on the effects of CR pressure and wave heating on the gas. Other effects such as thermal conduction and cooling mechanisms are not considered here. The result is a rapid ($\sim$ 100 Myr) destruction of the cloud as shown in Figure \ref{case3dens}.

In addition to the destruction of the cloud, the cloud's position varies with time. The initial onset of CRs causes a total pressure gradient which pushes the cloud to the right. This is a physical dynamical effect we will explore in \S\ref{sec:regimes}. But the combination of 1D planar symmetry and a reflective boundary condition on the left edge of the domain means that the gas to the left of the cloud becomes more rarefied as the cloud moves to the right. So eventually the pressure gradient evens out and then inverts, pulling the cloud back to the left (see discussion and footnote in \S\ref{sec:regimes}). We consider this latter dynamic an artifact of our 1D setup which should vanish in higher-dimensional simulations. We intend to explore this in future work.

This simulation exhibits some additional interesting properties. Figure \ref{shock} shows the different pressures at the same times plotted in \ref{case3dens}. The dot-dashed lines at the bottom display the CR pressure, the dashed lines above them are the gas pressure, and the solid lines are the total (CR + gas) pressure. We can see a sound wave propagating outward -- this is created by the impact of the first source-produced CRs on the outward-facing edge of the cloud. We found that despite the outflow conditions at the outer boundary, the wave eventually bounced back,  interfering with the simulation. To avoid this we simply placed the outer boundary far away from the cloud, using a non-uniform spatial grid for ease of computation. The test case shown here uses a grid that spans about 40 kpc, with 400 cells. The width of each cell is larger than the one preceding it by a constant factor, in this case 1.014. This results in grid widths of 2.2 pc on the left end, increasing to 16 pc at $z=$ 1 kpc, and to 580 pc at the far right end.

There is also a small feature leading the sound wave. This is a ``streaming front,'' which traces the position of the first source-emitted CRs as they stream outward. Whether this feature is resolved well or not depends on the smoothing scale $\epsilon$ in equation \eqref{smoothing}. We note that for this simple test we use very modest resolutions to reduce computation time. Because of this, neither the streaming front nor the sound wave front are well-resolved, but these are not the focus of this test.

\begin{figure}
\includegraphics[width=7cm,trim=1cm 5cm 4cm 0cm]{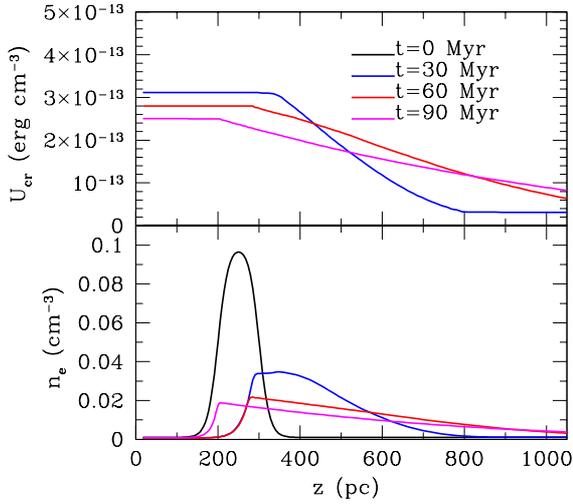}
\caption{Evolution of the gas density from an initial cool cloud due to the presence of a CR source at the left-hand boundary. Radiative cooling and thermal conduction are not included in this test. The cloud is quickly destroyed.}\label{case3dens}
\end{figure}
\begin{figure}
\includegraphics[width=7cm,trim=1cm 8cm 4cm 0cm]{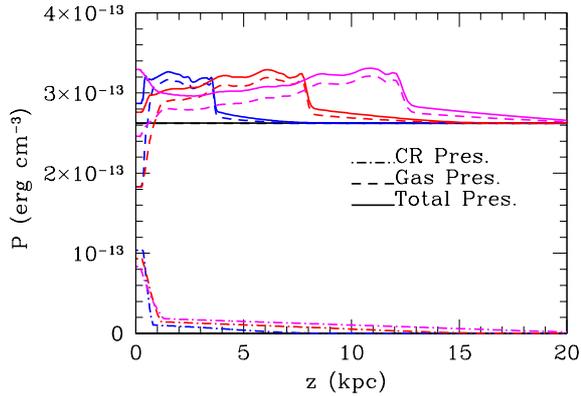}
\caption{Evolution of the CR pressure (dotted lines), gas pressure (dashed lines), and total pressure (solid lines) at the same times as in figure \ref{case3dens}. A sound wave propagates outward, led by the streaming CRs.}\label{shock}
\end{figure}
We can investigate the relative effects of CR pressure and wave heating by turning each off selectively (Fig.\ref{compare}). What we find is straightforward -- if we turn on wave heating but remove the effect of CR pressure on the gas (top panel), the cloud is destroyed, but does not move from its initial position. If we turn off wave heating but leave in the effect of CR pressure on the gas, the cloud is slowly pushed on, but not significantly distorted beyond a slight stretching (bottom panel). Note that in these runs we have placed the cloud farther away from the reflective boundary to minimize the effects of the 1D plane symmetry discussed above.

We emphasize here that, in the latter case, the pressure gradient from the onset of CR production is a transient effect. As discussed above, after some time the total pressure levels off and the cloud is no longer accelerated. The momentum transfer should then be thought of as a `kick' rather than a perpetual acceleration.

\begin{figure}
\includegraphics[width=7cm,trim=1cm 2cm 4cm 0cm]{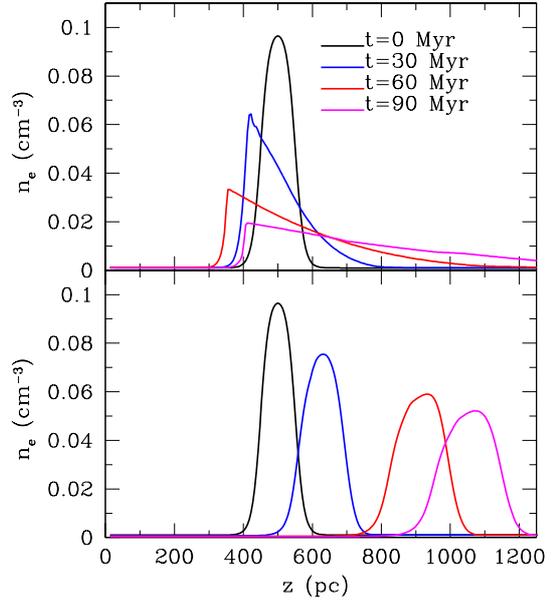}
\caption{Evolution of the gas density with different components turned off. The top plot has no CR pressure force, whereas the bottom plot has no wave heating.}\label{compare}
\end{figure}

\subsection{Imparting Momentum to Cool Clouds}\label{sec:regimes}
Ideally we would incorporate radiative cooling and thermal conduction into these 1D simulations. But the spatial resolution required to resolve the relevant length scales (see \S\ref{sec:lengthscales}) can be computationally expensive, particularly early in the evolution before equilibrium is reached. In the final state, the front can be quite broad (see Fig \ref{fig:length_scales}), mainly because of the pressurizing effects of CRs, which reduce gas densities and lower the rate of cooling. It is the transient which can be narrow.

We continue our analysis under the assumption that CR heating dominates over thermal conduction (see \S\ref{sec:lengthscales} for justification), which we therefore ignore. We also assume that a steady state is reached at the cloud interface where radiative cooling balances CR heating.

The effect of the CRs can then be broken down into two separate arenas: the CR pressure gradient imparts momentum to the cool cloud as a whole, while wave heating heats and thickens the cloud interface. The former effect can be studied with our simulations above so long as we turn off CR heating of the gas (the energy loss of CRs from this effect of course remains active). The latter effect can be studied in steady-state approximations, as we describe in section \ref{sec:steadystate}.

\begin{figure}
\includegraphics[width=6cm,trim=1cm 0cm 6cm 0cm]{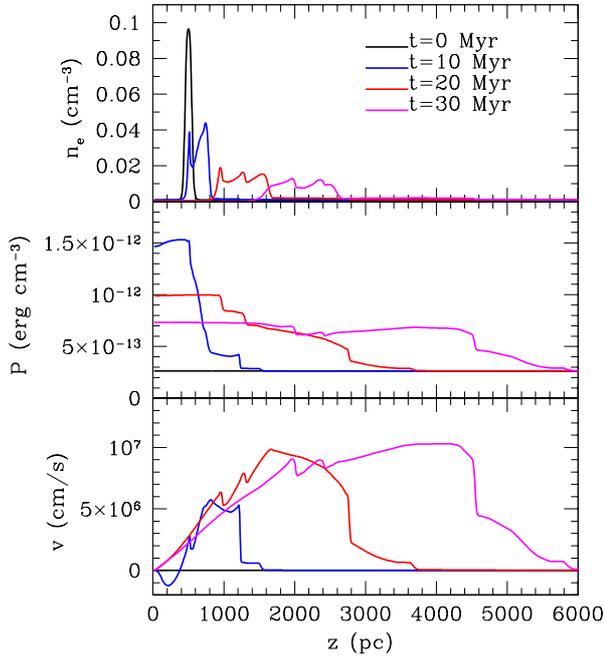}
\caption{Evolution of the gas density (top), total pressure (middle), and velocity (bottom) in the efficient cooling limit - the CR heating term in the gas energy equation is turned off, but CRs still lose their energy.}\label{fig:noheatfinal}
\end{figure}

\begin{figure}
\includegraphics[width=6cm,trim=1cm 11cm 6cm 2cm]{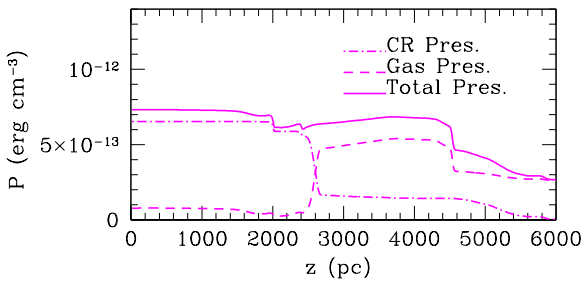}
\caption{Components of pressure at $t=30$ Myr. The CR pressure dominates inside and to the left of the cloud.}\label{fig:noheatpres}
\end{figure}

To consider again the time-dependent simulation without CR heating, see figure \ref{fig:noheatfinal}. This is the same setup as above in Figure \ref{compare} but the CR source is 20 times stronger in this simulation to obtain rough equipartition between the thermal gas and the CRs beyond the outward-facing edge of the cloud. Figure \ref{fig:noheatfinal} displays the cloud density, total pressure, and gas velocity at different times in the simulation.   The small nonmonotonic features in the profiles of density, velocity, and pressure seen in Figure \ref{fig:noheatfinal} are reproduced by three different numerical resolutions, as shown in the Appendix (Figure \ref{fig:res18}) at a time of 18 Myr, just short of the second time slice shown in Figure \ref{fig:noheatfinal}.

The dynamics of this situation are as follows. As the CRs first build up at the outward edge of the cold cloud, the resulting total pressure gradient (fig \ref{fig:noheatfinal} middle panel, blue line) pushes the cloud to the right. The cloud (as well as the hot material to its left) is stretched as it moves, gradually lowering the density (Figure \ref{fig:noheatfinal} top panel) and thermal pressure to the left of the CR bottleneck until total pressure equilibrium is re-established\footnote{In 1D, this rarefaction continues as the cloud progresses, eventually inverting the total pressure gradient and pulling back on the cloud like a spring. We do not expect this behavior in higher dimensions, as hot gas is available to flow in from directions perpendicular to the CR flow, preventing the pressure leftward of the cloud from dropping further. We therefore do not consider the results of these simulations to be robust at late times.}. As this happens, the material in the cloud is also accelerated to high velocities (fig \ref{fig:noheatfinal} bottom panel). We can then describe the large scale effect of the CRs on the cloud in two parts. The CRs have stretched the cloud, reducing the density inside. They have also imparted momentum to the cloud, accelerating it to some maximum velocity. Note that the resulting cloud velocity in this simulation ($\sim$ 100 km/s) is comparable to measurements of halo clouds in the Milky Way (\cite{wakker12}).

Figure \ref{fig:noheatpres} shows the different pressure components ($P_c$, $P_g$, and $P_{tot}$) at 30 Myr. At this time, total pressure equilibrium has been roughly reestablished. To the left of the cloud's outward edge (near 2600 pc), CR pressure dominates, the thermal pressure having been reduced by the stretching effect. To the right, the gas pressure is more significant. The transition between these two regions is exactly the CR bottleneck effect, and will be examined further in section \ref{sec:steadystate}. Far to the right, the sound wave generated by the CRs initial impact on the cloud travels outward (near 4500 pc) at the sound speed. Beyond that (near 6000 pc) is the leading edge of the first CRs generated at the beginning of the simulation, traveling outward at the Alfv\'en speed.

\subsection{Effects of Cloud Edge Width}
We may wonder if changing the initial thickness of the cloud edge has an appreciable affect on the cloud's dynamics. A thinner cloud edge would induce a steeper CR pressure gradient, which could in principle change how the cloud evolves. To test this we set up two clouds with different initial edge widths and compared the results at $t=30$ Myr.

Figure \ref{fig:pushcomp} shows the results of this comparison. After 30 Myr, the $\Delta z=25$ pc cloud is more spread out than the $\Delta z=10$ pc cloud. However, both clouds reach the same velocity, indicating that the width of the cloud edge does not affect the strength of the momentum kick the cloud receives from the CRs.

This makes sense. Overall, the force that a cloud receives depends on the total pressure drop between one side of the cloud and another, $F \sim (\Delta P) A$. As long as there are no dynamical or thermal instabilities, and Alfv\'enic/sound-crossing times are short, the lengthscale on which this drop takes place is irrelevant. However, thinner fronts are heated more intensely (see \S\ref{sec:steadystate}), which will also affect the cloud's final state.

\begin{figure}
\includegraphics[width=6cm,trim=1cm 5cm 6cm 0cm]{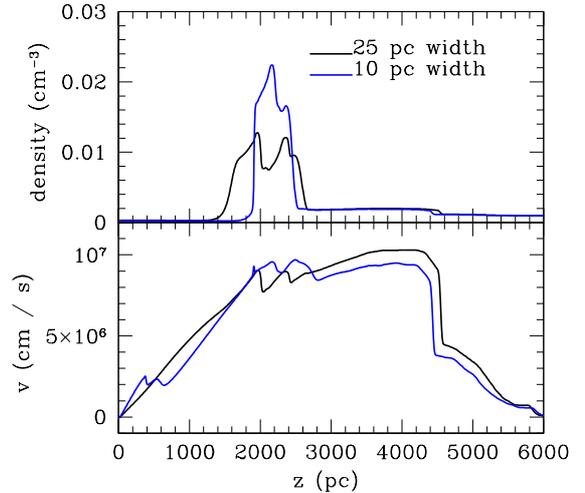}
\caption{Evolution of the gas density and velocity for clouds of different initial edge widths. All plots show the state of the cloud at $t=30$ Myr after the CR source is turned on.}\label{fig:pushcomp}
\end{figure}

\subsection{Are Clouds Important Energy Sinks?}

The preceding illustrative simulations make clear that CRs transfer energy and momentum to clouds. All of the energy and momentum which is usually transferred over large length scales in the hot medium is now deposited in a thin boundary layer. This raises the question of whether cold clouds can extract arbitrarily large amounts of energy from CRs, rendering CR dynamics unimportant downstream of the cloud. 

Let us consider a constant CR source which produces a constant flux $F_{0}$ of cosmic rays, and sends CRs streaming down a monotonically decreasing gas density gradient (assume that the B-field is constant). Now suppose we insert a warm ionized cloud in the middle. The CRs will be uncoupled until they exit the warm cloud. Thus, the problem is equivalent to inserting the source at the 2nd cloud interface, where the minimum in Alfv\'en speed arises. Since the cosmic rays are locked to the gas here, we have $P_{c} \propto v_A^{-\gamma_c}$, and $F \sim P_c v_A$. In this case, we have $F \propto v_{A}^{1-\gamma_{c}} \propto v_{A}^{-1/3} \propto \rho^{1/6}$. Thus, the CR flux does decrease as it propagates down a density gradient and heats the gas, but it does so very slowly. For instance, at the CRs propagate through a density contrast of $\sim 100$ (as at the interface of $T\sim 10^{4}$K and $T\sim 10^{6}$K gas), the CR flux decreases only by a factor of 2. The change in the CR flux when there is a cold cloud in the middle or not changes is only of order unity. 

This seems counterintuitive. CR heating causes the CRs to e-fold roughly every density scale-height. One should be able to able to arbitrarily sap energy from the CRs by increasing the number of e-foldings. However, the timescale on which this efolding happens is the Alfv\'en crossing time, which is now longer. So the rate at which the CRs lose energy depends only weakly on the density contrast: $\dot{E}_{c} \sim v_A \cdot \nabla P_{c} \sim v_{A} v_{A}^{-\gamma_{c}}/L \propto (\Delta \rho)^{1/6}$. 

Boundary conditions are important: the key is to consider a constant CR source. If one assumes, for instance, CRs in equipartition with thermal gas at the starting point, then the CR energy density does decrease rapidly.

This thought experiment also brings out the possibly episodic nature of cosmic ray forcing, even in the presence of a steady source - the more so if we consider multiple clouds lying along the same magnetic flux tube. The bottleneck effect is set up on a timescale of order the Alfv\'en travel time between the source and the point at which $v_A$ reaches its minimum. Prior to forming the bottleneck, the cosmic ray pressure gradient acts on the ambient gas to accelerate and heat it; afterwards it does not. Likewise, if cosmic rays encounter a second cloud after streaming through the first, a second bottleneck will be set up, affecting the evolution of the first cloud. Exploring the time averaged behavior of a system with multiple clouds and time modulated sources is beyond the scope of this paper, but it is likely to be very rich.

\section{Steady State 1D Thermal Fronts}\label{sec:steadystate}
Having examined the large-scale effects of a CR pressure gradient, we now zoom in on the interface between the cold and hot gas phases to describe the small-scale effect of CR wave heating. Since this interface contains intermediate-temperature ($\sim$ $10^5$ K) gas, its theoretical structure is closely tied to observations of certain ion column densities (see \cite{wakker12} and Figure \ref{fig:ratioplot}), which provides an opportunity to test the model. Thus, our main goal in this section is to understand the effect of CRs on the interfaces (and resulting column density predictions) of cold clouds embedded in the hot medium of a galaxy. 

Inside these clouds we expect CRs to be uncoupled from the gas, either because of strong ion-neutral wave damping or, in the bottleneck situation we are considering, because CRs stream sub-Alfv\'enically and the streaming instability is not activated. We can therefore focus on the temperature profile at the outward edge of the cloud only, where the CRs exit and recouple to the gas.

There is an extensive literature on thermal fronts. The properties of thermal conduction mediated fronts between two thermally stable phases such as warm and cold neutral interstellar gas are reviewed by \citet{inoue06}. Generally such fronts are not static; if the ambient pressure $P$ is below a critical value $P_{crit}$ the cold gas evaporates, if $P >P_{crit}$ the warm gas condenses, and only if $P=P_{crit}$ do the front-integrated values of cooling and heating balance, resulting in a static front. The evaporation/condensation flows are generally highly subsonic: the fronts are nearly isobaric, and the enthalpy flux is much greater than the kinetic energy flux.

The characteristic thickness of conduction mediated thermal fronts is the Field length $\lambda_F$, the length scale over which the net cooling rate $\rho{\mathcal{L}}$ (see equation \eqref{eq:gas3}) balances the divergence of the conductive heat flux \citep{field65,begelman_global_1990}: 
\begin{equation}
\lambda_{\rm F} = \left(\frac{\kappa T}{\rho{\mathcal{L}}} \right)^{1/2} 
\label{eqn:field_length}
\end{equation} 
where $\kappa = \, 5.6 \times 10^{-7} T^{5/2} \, {\rm erg \, s^{-1} \, K^{-1} \, cm^{-1}}$ is the Spitzer conductivity. The  thermally unstable gas in the front is stabilized by conduction. The Field length is equally significant for bounded clouds. Cool clouds immersed in a hot medium can only undergo conductive evaporation if their size is less than $\lambda_F$ \citep{cowie77, mckee77, mckee90}.

Due to the diffusive nature of conduction, the structure of all steady state fronts mediated by classical thermal conduction is determined from a two point boundary value problem which can be cast as an eigenvalue problem for the rate at which mass is exchanged between the phases. In contrast, fronts heated by streaming cosmic rays locked to the gas are modeled as a lower order system, and can have a range of mass fluxes.

In \S\S\ref{sec:gas_equations} and \ref{sec:1Dfronts} we derive a 1D semi-analytic model of such a cosmic ray dominated interface in the steady state, where radiative losses are balanced by CR heating. In the examples presented here, the mass flux is set to zero, indicating a static interface. This greatly simplifies the evolution equations, with the caveat that time dependent effects, and the effects of mass flux on the front structure, should be systematically explored in the future. 

Because these cosmic ray dominated front models represent a limiting case, in \S\ref{sec:lengthscales} we derive the conditions under which thermal conduction can be neglected, and in \S\ref{sec:frame_lock} we do the same for cosmic ray diffusion. While both these processes are indeed subdominant in the examples presented here, they could be significant in other cases. We are also well aware that including thermal diffusivity, cosmic ray diffusivity, or both, requires additional boundary conditions and therefore changes the nature of the solutions, at least through the formation of boundary layers, but possibly also globally. These are topics for future work.

\subsection{CR and Gas Equations}\label{sec:gas_equations}
Our simple model consists of a 1D plane parallel symmetric profile of gas. The cold region, representing the inside of the cloud, is on the left, and the hot region outside of the cloud is on the right. The intermediate-temperature interface lies between these two regions, and we imagine a CR source on the far left. As long as the gradient in gas density is monotonic, we can use equation \eqref{CR1} to determine CR pressure (again, in the limit of negligible CR diffusion).

Since we are only considering the structure of the cloud interface our $v_A$ profile will be monotonic. We can therefore use equation \eqref{CR1} for our analysis here. To model the gas we use equations \eqref{eq:gas1} - \eqref{eq:gas3} with the partial time derivatives and gas velocity $u$ set to zero to emulate a static steady state. We do not model MHD here, and instead assume a uniform magnetic field perpendicular to the plane of the front. Finally, the equation of state for the gas gives us a complete set of equations to solve:
\begin{equation}\label{eq:CRlocked}
P_c v_A^{\gamma_c}=\tr{constant}\Rightarrow P_c=P_{c,0}\left(\frac{n}{n_0}\right)^{\gamma_c/2}
\end{equation}
\begin{equation}\label{eq:state}
P_g=nk_BT
\end{equation}
\begin{equation}\label{eq:gas5}
\dd{}{z}(P_g+P_c)=0
\end{equation}
\begin{equation}\label{eq:gas6}
0=-\rho\mathcal{L}-v_A\dd{P_c}{z}
\end{equation}
In the above, $P_{c,0}$ is the CR pressure at some reference density $n_0$, and $\gamma_c$ is taken to be 4/3.

Unlike our time-dependent ZEUS simulations, we now include the radiative loss terms that appear in equation \eqref{eq:gas6}. We use a radiative cooling rate of the form $\Lambda(T)n^2$, with cooling function $\Lambda(T)$ determined from assuming collisional ionization equilibrium (CIE) and solar metallicity \citep{gnat07}. We also include a supplemental heating term of the form $\Gamma n$ to establish the two equilibrium phases on either side of the interface. The net cooling, to be exactly balanced with CR wave heating, is then
\begin{equation}\label{eq:equi}
-\rho\mathcal{L}=n^2\Lambda(T)-n\Gamma
\end{equation}

Equations \eqref{eq:CRlocked} - \eqref{eq:equi} can be reduced to a single first-order equation in the gas density $n$ as follows. Taking the spatial derivative of equation \eqref{eq:CRlocked},
\begin{equation}
\dd{P_c}{z}=P_c\frac{\gamma_c}{2}\frac{1}{n}\dd{n}{z}.
\end{equation}
and substituting the result into equation \eqref{eq:gas6} yields
\[
0=n^2\Lambda(T)-n\Gamma-v_AP_c\frac{\gamma_c}{2}\frac{1}{n}\dd{n}{z}
\]
\begin{equation}\label{eq:fronteq}
\Rightarrow \dd{n}{z}=n\frac{2}{\gamma_c}\left(\frac{n^2\Lambda(T)-n\Gamma}{v_AP_c}\right)
\end{equation}
where we use the remaining two equations to determine $T$ as a function of $n$:
\[
P_g+P_c=P_\tr{tot}=\tr{constant}
\]
\begin{equation}
\Rightarrow nk_BT=P_\tr{tot}-P_{c,0}\left(\frac{n}{n_0}\right)^{\gamma_c/2}.
\label{eqn:pressure_balance} 
\end{equation}

Note that CR pressure can significantly alter the density profile of the front. It becomes increasingly dominant in low temperature gas, significantly reducing its density. We can characterize this. For cosmic rays to be dynamically important in the hot gas (i.e., to drive a CR wind), it must be of comparable energy density to the thermal component, $X_{\rm CR}^{\rm hot} \equiv (P_c/P_g)^{\rm hot} \sim \mathcal{O}(1)$. Equation \eqref{eqn:pressure_balance} can be rewritten in the form:
\begin{equation}
\frac{n^{\rm cold}}{n^{\rm hot}} = \left( 1+ \frac{1}{X_{\rm CR}^{\rm hot}} - \frac{P_g^{\rm cold}}{P_g^{\rm hot}} \right)^{3/2} \approx \left( 1+ \frac{1}{X_{\rm CR}^{\rm hot}} \right)^{3/2}
\label{eqn:overdensity} 
\end{equation}
where the last approximation holds for $X_{\rm CR}^{\rm hot} > 0.1$, which implies $X_{\rm CR}^{\rm cold} > 10$, and $P_g^{\rm cold}/P_g^{\rm hot} < 0.1$. Since the effects of CR pressure increase rapidly at low temperatures (higher densities), the density contrast is much smaller than one might expect if one only assumes thermal pressure balance. 

\subsection{Numerical 1D Fronts}\label{sec:1Dfronts}
We obtain numerical solutions to \eqref{eq:fronteq} with a simple forward integration. These front solutions are characterized by three parameters: the gas pressure in the cold region $P_g^{\rm cold}$, the ratio of CR pressure to gas pressure in the cold region $X_{\rm CR}^{\rm cold}$, and the supplemental heating coefficient $\Gamma$. Note that in this framework, increasing the parameter $X_{\rm CR}^{\rm cold}$ also increases the total pressure $P_\tr{tot}$. All models use a fixed B-field of 3 $\mu$G.

The pressures for two example fronts are shown in Figure \ref{fig:fronts}. Both models use $P_g^{\rm cold}=1000k_b$ and $\Gamma=1\ee{-26}$ erg s$^{-1}$. The top plot uses $X_{\rm CR}^{\rm cold}=2$, the bottom $X_{\rm CR}^{\rm cold}=10$. Note that these parameters correspond to relatively modest CR relative energy densities of $X_{\rm CR}^{\rm hot}=4.6\%, 13\%$ respectively in the hot gas, at the low end of typical parameters for CR-driven winds. Thus, CR-driven winds inevitably imply that cold gas interfaces will be highly pressurized by cosmic rays, and we should not expect thermal pressure balance.

\begin{figure}
\includegraphics[width=9cm]{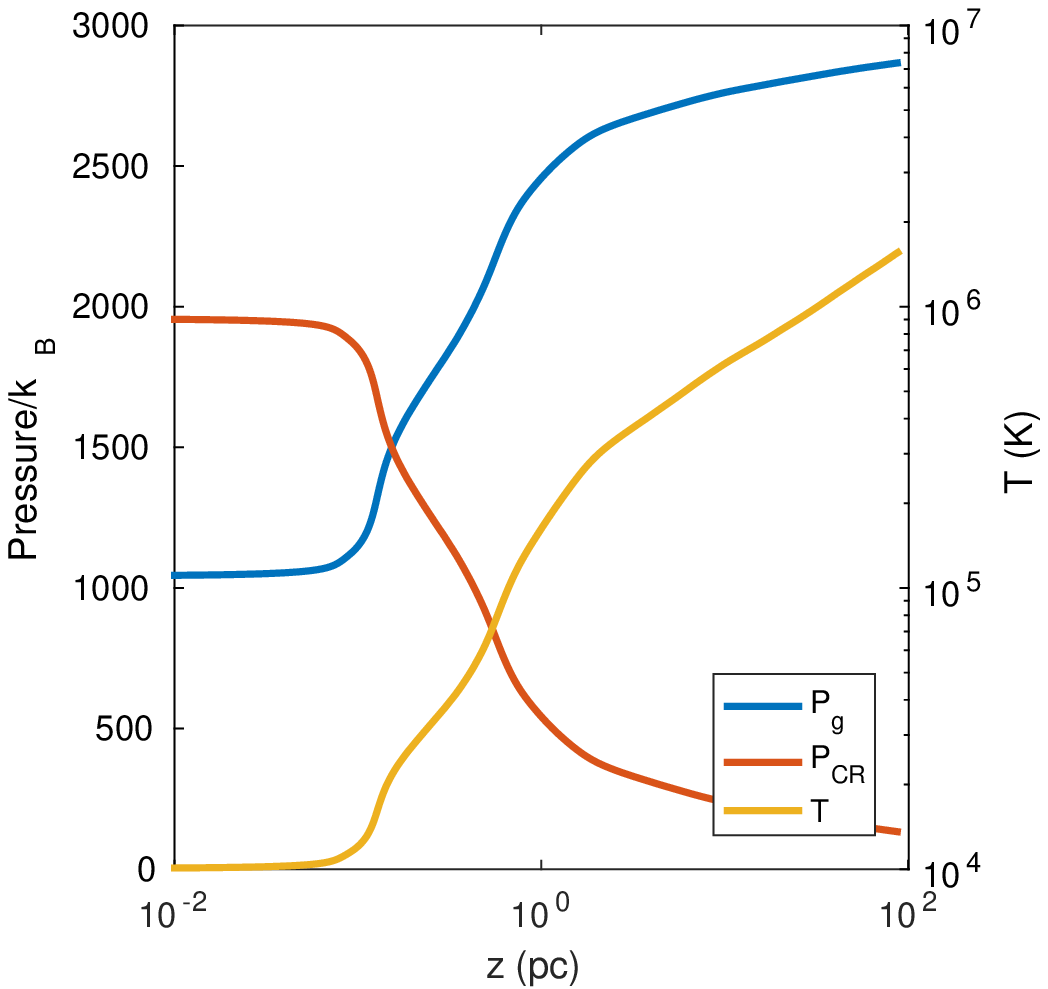}
\includegraphics[width=9cm]{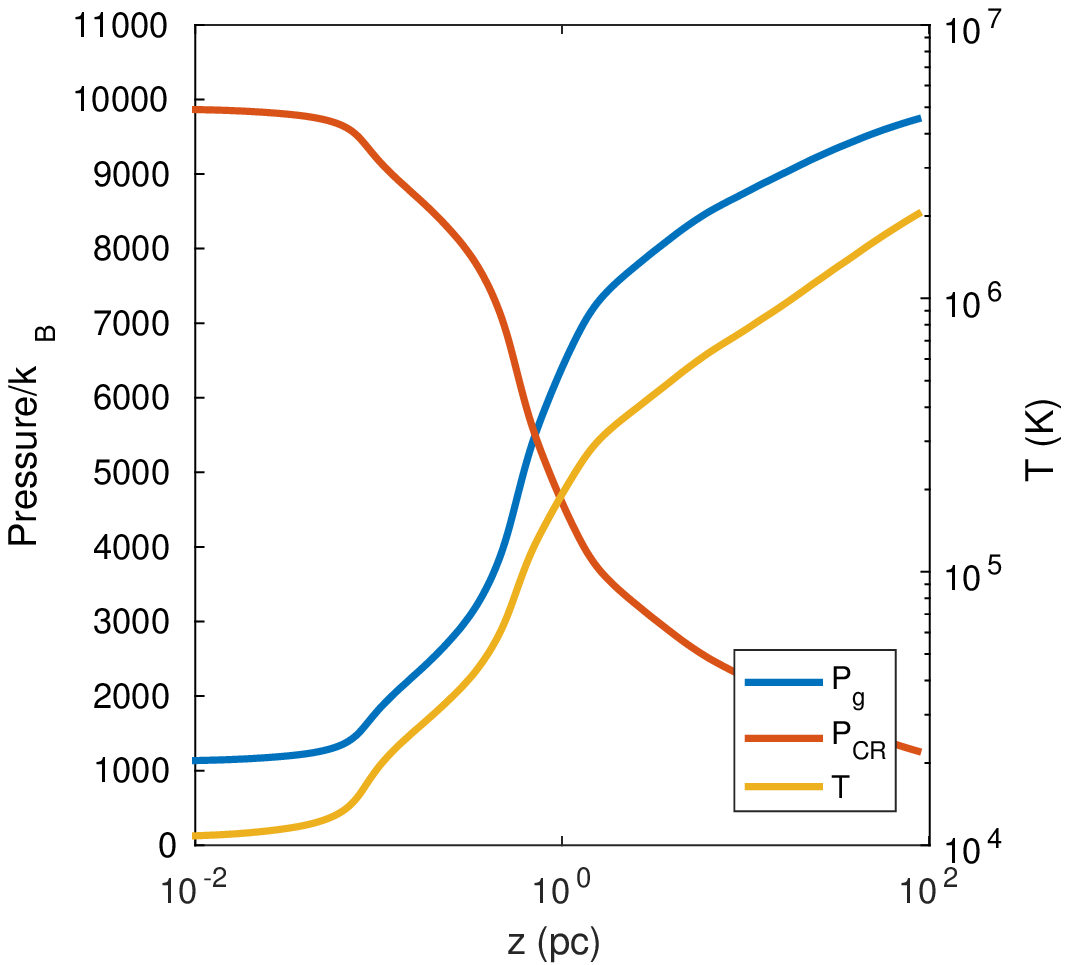}
\caption{$P_c$ and $P_g$ for two 1D front models with different values of $X_{\rm CR}^{\rm cold}$. Top: $X_{\rm CR}^{\rm cold}=2$. Bottom: $X_{\rm CR}^{\rm cold}=10$.}\label{fig:fronts}
\end{figure}

Using ionization fractions $X_i(T)$ from \cite{gnat07} and relative abundances $A_X$ from \cite{asplund09} we can calculate the column densities of ions defined by $N_X=A_X\int X_i(T) n \tr{d}x$, where $n$ is the gas density and the integral is across the front. To avoid dependence on the length of integration we use, we limit our integral to temperatures below some critical temperature $T_\tr{crit}$, for which we choose the temperature where the OVI ionization fraction drops to $\sim$0.001, or $T_\tr{crit}=2\ee{6}\ K$.

We calculate column densities for CIV, OVI, and SIV for each solution, covering a variety of values of $P_\tr{tot}$ and $X_{CR}$. The resulting ratios for a sample of models are shown in the top plot of Figure \ref{fig:ratioplot} alongside line ratio data from observations of high velocity clods (HVCs) in the Milky Way from \cite{wakker12} and some other select interface models. We zoom in on the CR front portion of the plot to see the dependence of the line ratios and column densities in the bottom plot of Figure \ref{fig:ratioplot}.

\begin{figure}
\includegraphics[width=9cm]{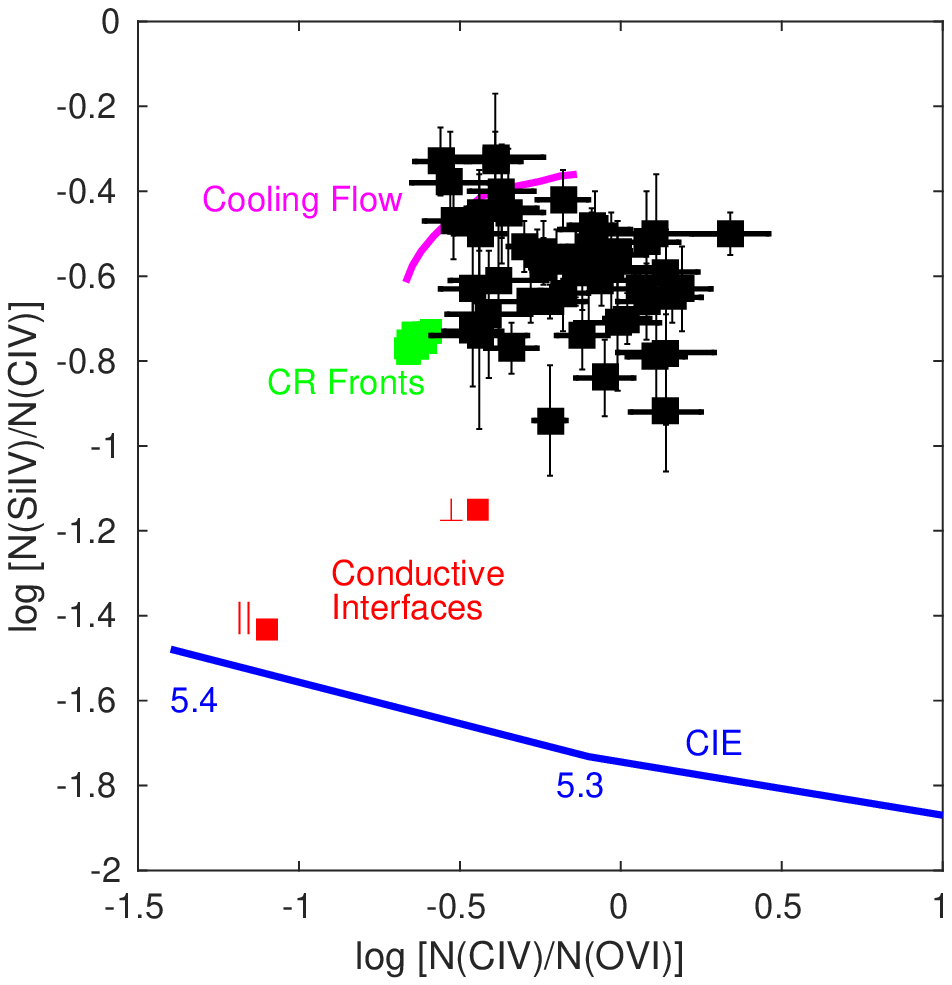}
\includegraphics[width=9cm]{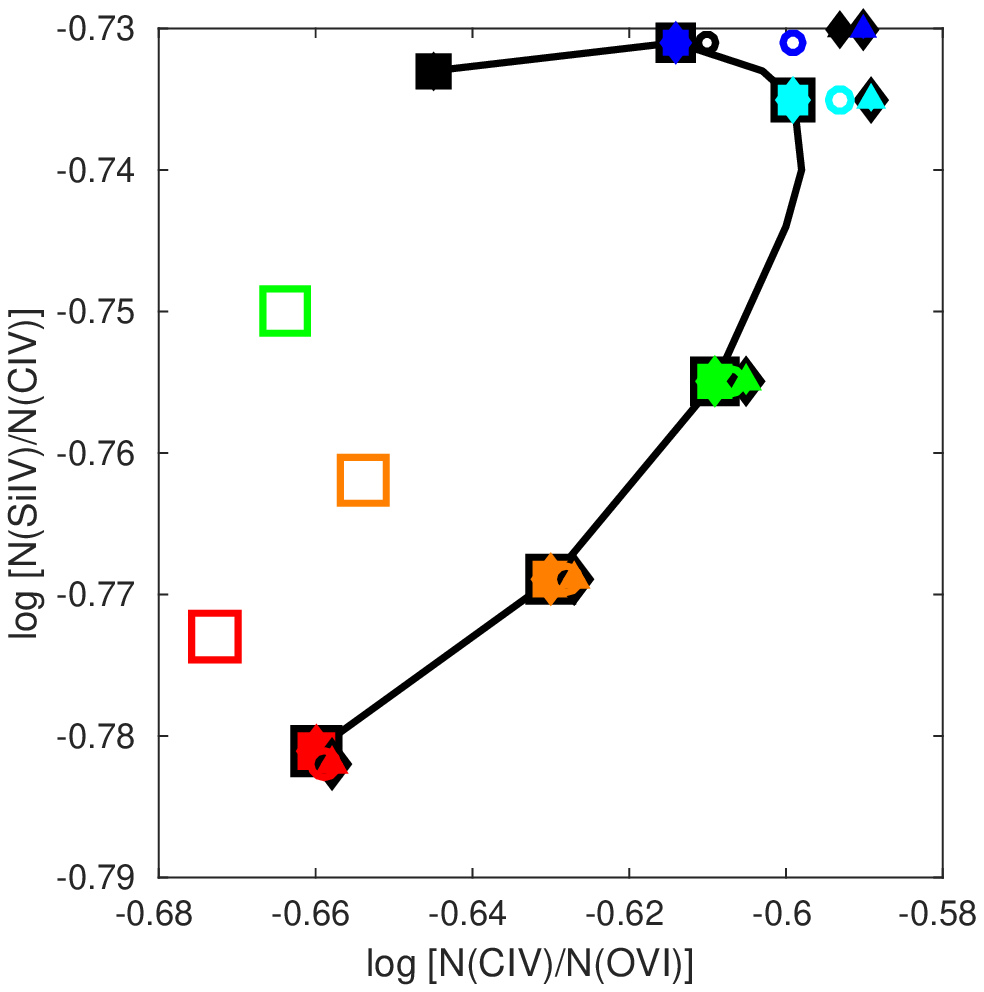}
\caption{Top: Predicted line ratios for a selection of models. The 1D plane parallel models with CR heating introduced in this paper are shown in green and labelled ``CR Fronts''. Also shown are data for line ratios seen in Galactic clouds. Bottom: Closeup of different CR front models (note the different $x$ and $y$ axes). Model parameters are indicated by symbol shape ($P_g^{\rm cold}$), color ($X_{\rm CR}^{\rm cold}$), and symbol filling ($\Gamma$) (see table \ref{tab:legend}). Predicted observations are indicated by position (ratios) and size (SiIV column density) of the points.}\label{fig:ratioplot}
\end{figure}

Consider first the top plot, which compares our CR front models to some other models, as well as to observations. Our model is a good fit to the data, and is competitive with other models. Our predictions of the CIV to OVI ratio are on the low side, but this might be remedied with photoionization, which we have ignored. Photoionization would boost the lower-potential SiIV and CIV ion densities, but would not appreciably increase the OVI density (see \cite{wakker12}, Appendix A). Including this effect would shift our predictions to the right on this plot. How much they would shift would require more analysis than we cover in this work.

\begin{table}
 \begin{center}
  \begin{tabular}{l l | l l | l l}
    \hline
    $X_{\rm CR}^{\rm cold}$ & Color & $P_g^{\rm cold}$/$k_b$ & Shape & $\Gamma$ & Filling \\
    & & [K cm$^{-3}$] & & [erg s$^{-1}$] \\
    \hline
    2 & Black & 100 & Square & 10$^{-25}$ & Open \\
    4 & Blue & 1000 & Diamond & 10$^{-26}$ & Filled \\
    10 & Cyan & 2000 & Circle \\
    40 & Green & 10000 & Triangle \\
    100 & Orange \\
    400 & Red \\
    \hline
  \end{tabular}
 \end{center}
\caption{Legend for Figure \ref{fig:ratioplot} bottom panel.}\label{tab:legend}
\end{table}

We can also compare the absolute ion column densities of our models to the data. To minimize the effect of our upper limit on temperature discussed above, we compare the SiIV column densities which are insensitive to this limit. \cite{wakker12} observe total SiIV column densities of log[N(SiIV)] = 13.57$\pm$0.18. The predicted SiIV column densities in for the models shown in figure \ref{fig:ratioplot} range from log[N(SiIV)] = 11.04 - 12.47, although we note that the highest of these come from unphysically thick ($\sim$kpc) interfaces. We find that column densities increase with increasing $X_\tr{CR}^\tr{cold}$ and with decreasing $P_g^\tr{cold}$. They are nearly unchanged for our two different choices of $\Gamma$. We also note again that low ionization lines like SiIV may be highly affected by photoionization, which these steady-state models do not account for.

We can alternatively compare OVI column densities. OVI is a high ionization line, and so is less affected by photoionization. \cite{wakker12} observes total OVI column densities of log[N(OVI)] = 14.21$\pm$0.20. Our models have OVI column densities ranging from log[N(OVI)] = 12.36 - 13.92. We note that our OVI predictions have their own issues. Thermal conduction is not accounted for, and will have a bigger impact at higher temperatures (see \S\ref{sec:lengthscales}). We have also implemented an integration limit of $T=2\ee{6}$ K - OVI has non-zero abundance up to about $5\ee{6}$ K, so if our high temperature equilibrium is less than this, our OVI column is technically infinite. Using an integration limit prevents this, but then our exact OVI column depends slightly on our choice of this limit. We chose $2\ee{6}$ K because that is the temperature where the OVI fraction drops to .001. Note that if densities are sufficiently low due to CR pressurization, then even high ions such as OVI can be affected by photoionization.

The column densities predicted by our models fall short of the observations, but this could simply be because we are comparing the column from a single interface to that of an entire sight line. If the average sight line passes through tens, or even hundreds of such interfaces (as expected if cold gas has a `fog'-like structure \citep{mccourt16}), the predicted column densities could match the observations. Also, as mentioned above, photoionization could significantly increase metal line column densities per interface.

The bottom panel of Figure \ref{fig:ratioplot} shows line ratios for CR fronts with different parameters. The primary factor for determining the line ratios seems to be $X_{\rm CR}^{\rm cold}$, indicated by the color of the points (see table \ref{tab:legend}). If this parameter is fixed, the line ratios are insensitive to the total pressure of the front. In contrast, the absolute column densities, indicated by the size of the points, are mainly determined by $P_g^{\rm cold}$ - lower gas pressures permit larger SiIV column densities. There is a small increase in column density with $X_{\rm CR}^{\rm cold}$ but the variation is small compared to the variation with $P_g^{\rm cold}$. Overall, the parameter dependence is relatively weak and CR heated fronts occupy a well-localized region in a line ratio plot, meaning that the model is quite testable. This arises because whenever CR heating is dominant, CR pressure support becomes important too, particularly at the lower temperatures associated with these ions ($X_{\rm CR} \propto n^{-1/3} T^{-1}$). In the limit where $X_{\rm CR}^{\rm cold} \gg 1$, then at low temperatures, the density $n(T)$ changes only weakly with temperature in the critical range $10^{4} \, {\rm K} < T < 10^{5} \, {\rm K}$ for these lines, independent of the exact value of $X_{\rm CR}^{\rm cold}$; thus column density ratios (though not the actual column densities) depend weakly on $X_{\rm CR}^{\rm cold}$. However, we caution once again that detailed predictions are likely to change significantly once photoionization is included.

Conversely, it is worth noting that CRs can significantly affect ionization parameter inferences, when photoionization dominates. As previously noted, CRs can significantly reduce cold gas densities (equation \eqref{eqn:overdensity}). Indeed, when this happens, the gas should move from the collisionally dominated regime to the photoionization dominated regime. Note that (with some notable exceptions, e.g. \citet{stern16}) photoionization modeling is generally done in the single zone regime, where all gas is photoionized at a single density. This is clearly not true for an interface with only thermal pressure, but becomes increasingly justified as $X_{\rm CR}$ increases. Interestingly, \citet{werk14} find from photoionization modeling of HST COS observations of cool, photoionized gas in the circumgalactic medium of 44 $L \sim L_{*}$ galaxies at $z\sim 0.2$ that cool gas densities are roughly two orders of magnitude lower than what one might expect from thermal pressure balance. Instead, the inferred gas densities are comparable to expectations for hot gas densities. This otherwise puzzling observation is compatible with equation \eqref{eqn:overdensity}, where $n^{\rm cold} \sim n^{\rm hot}$ for $X_{\rm CR}^{\rm hot} \sim \mathcal{O}(1)$.

\subsection{Characteristic Lengthscales}
\label{sec:lengthscales} 

When the CRs are locked to the fluid, $P_c \propto v_A^{-\gamma_c}$, there is a close (but not exact) analogy between thermal conduction and the effects of cosmic rays. Thermal conduction seeks to erase gradients in temperature, while CR coupling seeks to erase gradients in cosmic rays pressure $P_c$. In the limit of tight coupling, this is equivalent to gradients in Alfv\'en speed $v_A$; if the B-field is not strongly dependent on density, this thus seeks to erase gradients in density and thus (under isobaric conditions) temperature as well. 

We defined the Field length $\lambda_F$, the scale below which thermal conduction suppresses thermal instability, in equation \eqref{eqn:field_length}. Scaling the units to our problem
we can write
\begin{equation}
\lambda_{\rm F}  = 790 \, {\rm pc} \left( \frac{P_g}{10^{3} {\rm K \, cm^{-3}}} \right)^{-1}  T_{6}^{11/4} \Lambda_{-22}^{-1/2}.  
\label{eqn:num_field_length}
\end{equation} 

Similarly, in a CR heating dominated front, the characteristic temperature scale height from the energy equation is: 
\begin{equation}
\lambda_{\rm CR} =  \frac{v_A P_c}{n^{2} \Lscript} = X_{\rm CR} v_A t_{\rm cool}. 
\label{eqn:lambda_CR} 
\end{equation} 

We also speculate (but do not show) that in analogy to the Field length, only perturbations with $\lambda > \lambda_{\rm CR}$ can undergo isobaric thermal instability, in a background where CR heating balances cooling. Completing this analogy also suggests that cold clouds with $\lambda > \lambda_{\rm CR}$ can accrete and grow in mass, while cold clouds with $\lambda < \lambda_{\rm CR}$ evaporate. We will consider this problem in detail in future work. A full linear stability analysis is beyond this scope of this paper, and likely non-trivial\footnote{Analyses which ignore the vector nature of perturbations and approximate CR heating as a function of gas density suggest regimes where CR heating is stable \citep{loewenstein91, wiener13, pfrommer13}. \citet{sharma10} perform a vector analysis, but assuming that CR fluid is adiabatic.}. However, we do know that acoustic waves can be destabilized by cosmic ray pressure gradients \citep{drury_falle86,kang_etal92}, and that cosmic ray diffusion can itself be destabilizing \citep{begelman_zweibel94}, so the full problem is likely to be very rich.

In general, we can use equations \eqref{eqn:field_length} and \eqref{eqn:lambda_CR} to estimate the relative importance of thermal conduction and CR wave heating: if $\lambda_{\rm F} > \lambda_{\rm CR}$, then conduction dominates, and vice versa. However, we can get a useful bound from analytical considerations. The CR flux is $F_c \sim P_c v_A$, while the {\it maximum} conductive flux is given by the saturated heat flux $F_{\rm cond,sat} \sim P_g c_{\rm s}$ (see \S\ref{sec:frame_lock}). Thus, CRs are bound to dominate when $F_{\rm CR}/F_{\rm cond,sat} > 1$, or when $X_{\rm CR} > \beta^{1/2}$. In the $\beta \sim 1$ conditions of the ISM, CR heating dominates whenever CR and thermal pressure support become comparable.

\begin{figure}
\includegraphics[width=0.5\textwidth]{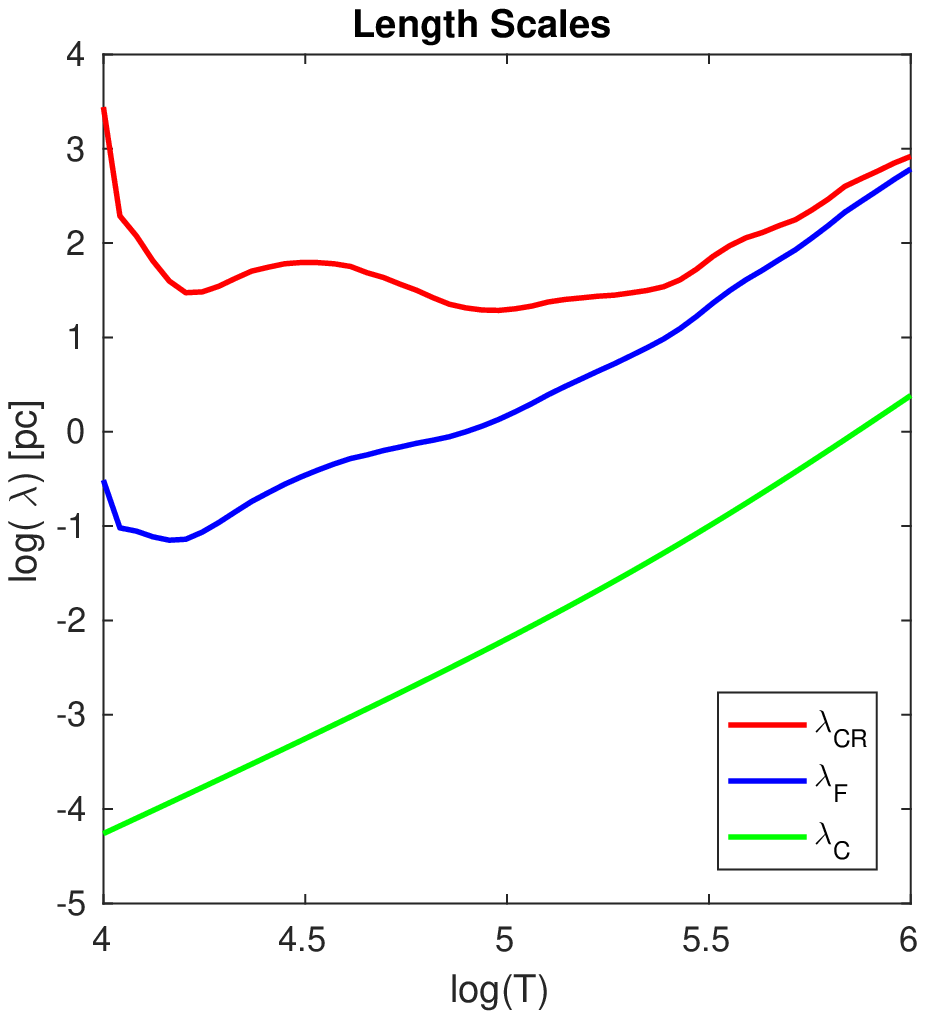}
\includegraphics[width=0.5\textwidth]{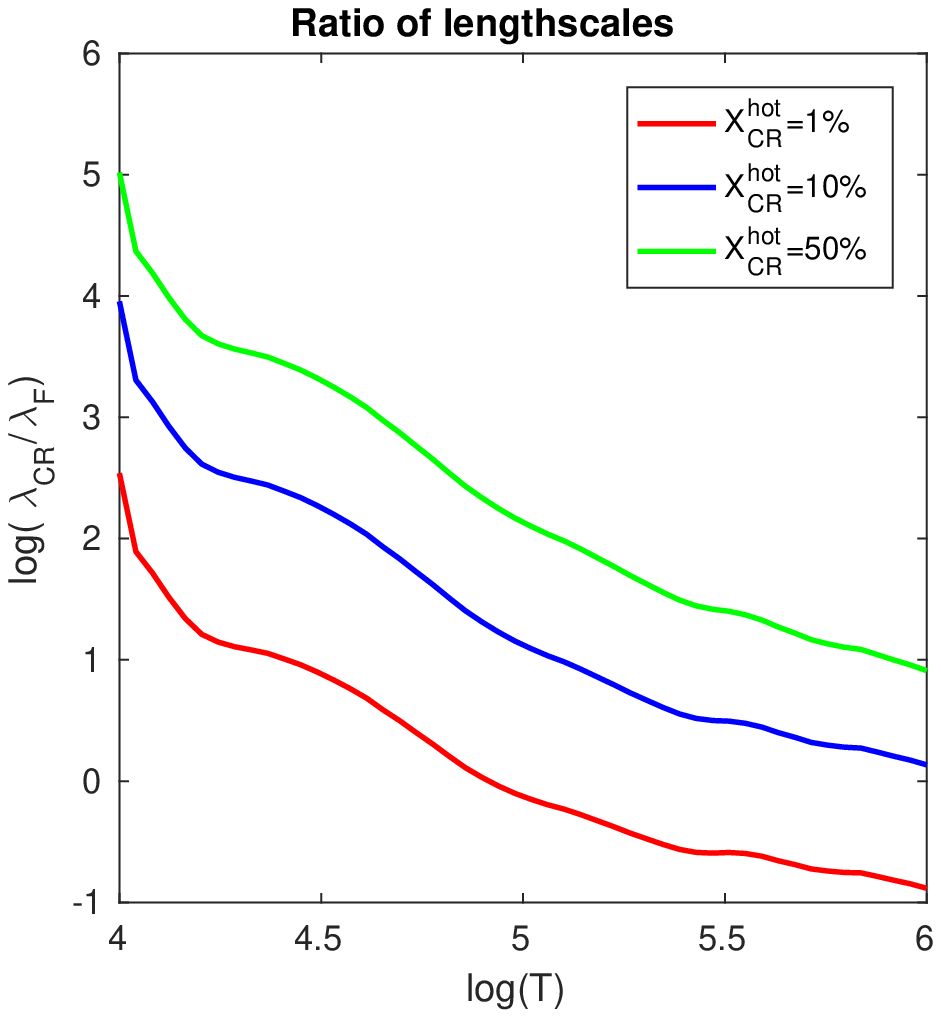}
\caption{Top panel: the CR scale height $\lambda_{\rm CR}$ (equation \eqref{eqn:lambda_CR}), Field length $\lambda_{\rm F}$ (equation \eqref{eqn:field_length}), and Coulomb mean free path $\lambda_{\rm C}$ (equation \eqref{eqn:lambda_C}) for $T^{\rm hot}=10^{6}$K, $n^{\rm hot}=10^{-3} \, {\rm cm^{-3}}$, $X_{\rm CR}^{\rm hot} = 0.1$. Bottom panel: the ratio $\lambda_{\rm CR}/\lambda_{\rm F} $ for $X_{\rm CR}^{\rm hot} = 1\%, 10\%, 50\%$, with all other parameters as above.}
\label{fig:length_scales}
\end{figure}

In the top panel of Fig \ref{fig:length_scales}, we show the CR scale height $\lambda_{\rm CR}$ (equation \eqref{eqn:lambda_CR}), Field length $\lambda_{\rm F}$ (equation \eqref{eqn:field_length}), and Coulomb mean free path
\begin{equation}
l_{\rm C} = \left(\frac{3 k T_{e}}{m_{e}} \right)^{1/2} t_{\rm ee} = 3 \, {\rm pc} \, T_{6}^{2} n_{-3}^{-1} 
\label{eqn:lambda_C}
\end{equation}
for $T^{\rm hot}=10^{6}$K, $n^{\rm hot}=10^{-3} \, {\rm cm^{-3}}$, $X_{\rm CR}^{\rm hot} = 0.1$. As expected, CR heating clearly dominates at low temperatures. Due to frame locking $P_c \propto \rho^{\gamma_{c}/2}$, the relative CR energy density $X_{\rm CR}$ rises at low temperatures; by contrast, the conductive flux falls rapidly. In the bottom panel, we show the effect of the CR pressure fraction in the hot gas $X_{\rm CR}^{\rm hot}$. The impact of CRs rises swiftly as $X_{\rm CR}^{\rm hot}$ increases, both because the CR heating rate increases and the radiative cooling rate decreases (due to CR pressurization reducing gas densities). If we combine equations \eqref{eqn:lambda_CR} and \eqref{eqn:overdensity}, then at low temperatures $\lambda_{\rm CR} \propto (X_{\rm CR}^{\rm hot})^{11/4}$.

From Fig. \ref{fig:length_scales}, we see that thermal conduction dominates at high temperatures, while CR heating dominates at low temperatures. For the observable line ratios we are interested in and likely values of $X_{\rm CR}^{\rm hot}$, CR heating dominates (the highest ion is OVI, which peaks at $T \sim 2 \times 10^{5}$K). Conductive heating at high temperatures is likely to be more important for controlling the overall mass flux between the hot and cold phases.

\subsection{Validity of CR Hydrodynamics and Frame Locking}
\label{sec:frame_lock}

Due to the small expected lengthscales in a CR heated front, it is worth checking our assumptions of CR hydrodynamics (i.e., equation \eqref{CR1}) and that CRs are locked to the Alfv\'en wave frame (equation \eqref{CR2}). The two are of course closely related; both rely on sufficiently rapid scattering of CRs by magnetic irregularities generated by the cosmic rays themselves. 

Cosmic ray hydrodynamics is valid if the CR mean free path $l_{\rm CR}$ is less than the CR pressure scale height $L_{\rm CR}$
\begin{equation}\label{eq:LCR}
L_{\rm CR} = \frac{P_c}{\vert\nabla P_c\vert} 
\end{equation}
while cosmic ray diffusion can be neglected if the advective flux $v_AP_c$ is the dominant term in the cosmic ray energy flux $F_c$ defined below equation \eqref{CR1}. Introducing the drift speed $v_D$ 
\begin{equation}\label{eq:vD}
v_D \equiv v_A +\frac{\kappa_c}{L_{\rm CR}},
\end{equation}
the cosmic rays are well locked to the wave frame if
\begin{equation}\label{eq:lock_parameter}
\frac{v_D}{v_A}-1 = \frac{\kappa_c}{v_AL_{\rm CR}}=\frac{l_{\rm CR}}{L_{\rm CR}}\frac{c}{3v_A}\ll 1,
\end{equation}
where we have used the usual relationship between diffusivity and mean free path. Equation \eqref{eq:lock_parameter} shows that if the CRs are well locked to the waves, the hydrodynamic approximation is valid, but the converse is not necessarily true.
 
The mean free path is related to the scattering fluctuation amplitude by \citep{kulsrud05}
\begin{equation}\label{eq:db}
l_{\rm CR}=\frac{c}{c_{\nu}\omega_p(\delta B/B)^2},
\end{equation}
where $c_{\nu} = \pi/2$ for cosmic rays with Lorentz factor of order unity, and $\omega_{\rm p}$ is the CR gyro-frequency.
 
The amplitude of the Alfv\'enic fluctuations is determined by the balance between wave driving and wave damping. In ionized gas, damping can arise from either (i) turbulent wave damping, arising from the non-linear interaction between Alfv\'en wave packets, which causes them to cascade to smaller scale, thus destroying them \citep{yan02,farmer04}, or (ii) non-linear Landau damping \citep{lee73,cesarsky81}, due to wave-particle resonance between thermal ions and the beat waves created by superposition of Alfv\'en waves. In this paper, we shall consider non-linear Landau damping (NLLD) exclusively. Firstly, on the small ($l \sim 0.01-0.1$pc) scales we consider, turbulent motions can be highly sub-Alfv\'enic and stabilized by magnetic tension. Secondly, if turbulent motions {\it are} important, then the physics of the front is likely to be strongly affected by turbulent mixing between the hot and cold phases \citep{begelman90,slavin93,kwak10}, and it is inconsistent to consider CR heating in isolation. 

If NLLD is the primary damping mechanism, then the Alfv\'en fluctuation amplitude $\delta B/B$, degree of super-Alfv\'enic drift $v_D/v_A$, and scale height $L_{\rm CR}$ are all related and can be determined self consistently \citep{loewenstein91,wiener13}. Here we follow \cite{wiener13}, except that we calculate these quantities for the bulk cosmic ray distribution, instead of as a function of energy, as was done in that paper, and we use equation \eqref{eq:vD} for $v_D$.  We make the further assumption that cosmic ray heating balances radiative cooling, and use equation \eqref{eq:gas6}. After some algebra, we find
\begin{equation}\label{eq:vDvA}
\frac{v_D}{v_A}-1=\left(\frac{1}{9}\sqrt{\frac{2}{\pi^3}}\frac{\alpha - 2}{\alpha - 3}\frac{c v_i}{v_A^2}\frac{n_i^3\Lambda m_pc^2\langle\gamma\rangle}{\omega_{cp}P_c^2}\right)^{1/2}.
\end{equation}
In equation \eqref{eq:vDvA}, the numerical prefactors come from the expressions for the cosmic ray streaming instability growth rate, NLLD rate, and scattering rate, $\alpha$ is the slope of the CR momentum spectrum, $v_i\equiv\sqrt{k_BT/m_i}$, and $\langle\gamma\rangle$ is the mean CR Lorentz factor. 

Equation \eqref{eq:vDvA} can be evaluated
numerically for parameters relevant to galactic halo clouds as
\begin{equation}\label{eq:vDvA2}
\frac{v_D}{v_A}-1 = 400\frac{T_4^{1/4}n_i^2\Lambda_{-22}^{1/2}\langle\gamma\rangle^{1/2}\mu^{1/2}}{B_{\mu G}^{3/2}P_{c,eV}}.
\end{equation}
In equation \eqref{eq:vDvA2} we follow the usual convention that for any quantity $Q$, $Q_x$ means $Q/10^x$ in cgs units and $B$ and $P_c$ are expressed in units of $\mu$G and eV/cm$^3$,
respectively. We also introduce the mean molecular weight $\mu$, which enters from the substitution of $v_A=B/\sqrt{4\pi\mu m_p n_i}$. The CR mean free path \eqref{eq:db} can be evaluated similarly:
\begin{equation}\label{eq:mfp}
\frac{l_\tr{CR}}{L_\tr{CR}}=\frac{3v_A}{c}\left(\frac{v_D}{v_A}-1\right)=8.7\ee{-3}\frac{n_i^{3/2}T_4^{1/4}\Lambda_{-22}^{1/2}\langle\gamma\rangle^{1/2}}{B_{\mu G}^{1/2}P_{c,eV}}
\end{equation}
This quantity is plotted in the top panel of figure \ref{fig:vdrift} for the front solutions shown in figure \ref{fig:fronts} and shows that the fluid approximation is robust for these particular two models (recall that these two models both have $P_g^{\rm cold}=1000k_b$ and $\Gamma=1\ee{-26}$ erg s$^{-1}$). We find that the fluid approximation is well-satisfied for all of the models presented in table \ref{tab:legend}.

The right hand side of equation \eqref{eq:vDvA2} is plotted in the bottom panel of figure \ref{fig:vdrift}. Here as elsewhere $B_{\mu G} = 3$. We see that for the parameters we have chosen, $v_D/v_A$ is of order unity, and the wave-locking assumption is marginally valid, at least in the $X_{\rm CR}^{\rm cold}=10$ model. The peaks seen are largely due to the peaks in the cooling curve. Looking at the rest of our models, the CRs are well-locked everywhere for low $P_g^\tr{cold}$ or high $X_\tr{CR}^\tr{cold}$, but the coupling can be very weak at $P_g^\tr{cold}>1000k_B$ and $X_\tr{CR}^{cold}<40$, with $v_D/v_A$ reaching as high as 45 in some locations in the worst case. We speculate that in these cases, the effective wave heating rate is less, resulting in thinner fronts and lower column densities than predicted. We further speculate that, since coupling is generally weakest at low temperatures, column densities for the low ions SiIV and CIV would drop more than for high ions like OVI, resulting in column ratios slightly to the left of our predictions in figure \ref{fig:ratioplot}.

\begin{figure}
\includegraphics[width=9cm]{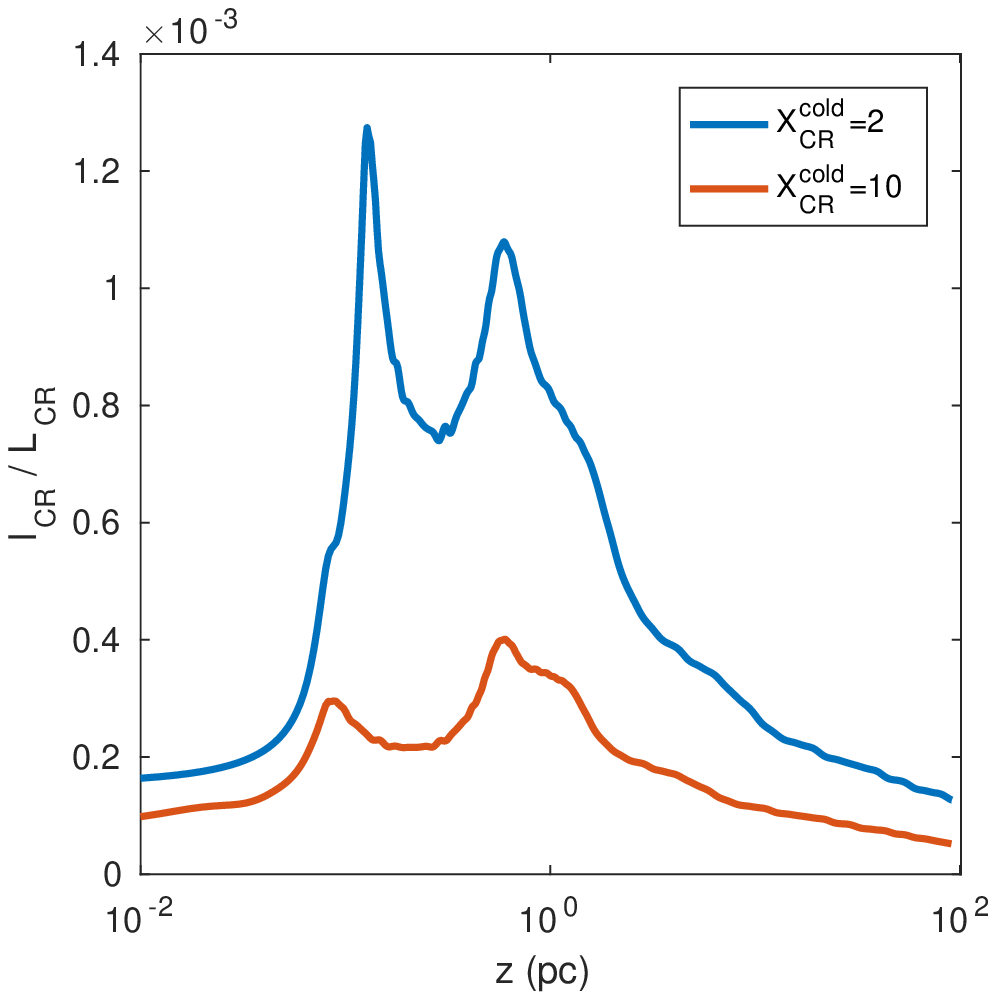}
\includegraphics[width=9cm]{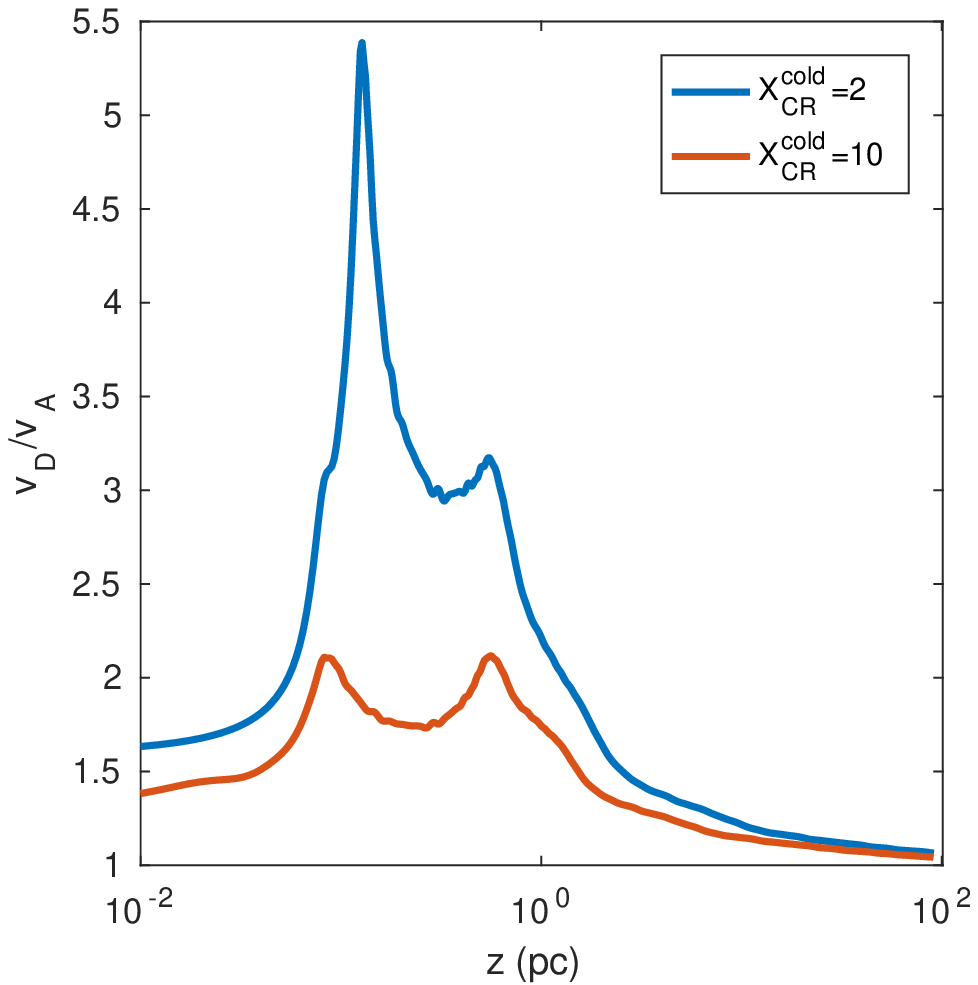}
\caption{Top: A posteriori evaluation of the CR mean free path (equation \eqref{eq:mfp}) to check the validity of our fluid approximation for the two models shown in figure \ref{fig:fronts}. The mean free path is much less than the CR scale height $L_\tr{CR}$ everywhere in both models. Bottom: A posteriori evaluation of the CR bulk drift speed (equation \eqref{eq:vDvA2}) to check the validity of our wave-locked assumption for the two models shown in figure \ref{fig:fronts}. The drift speed $v_D$ never exceeds the Alfv\'en speed by a factor of much more than 5 in the first model ($X_{\rm CR}^{\rm cold}=2$), or by 2 in the second model ($X_{\rm CR}^{\rm cold}=10$).}\label{fig:vdrift}
\end{figure}

Note that the relative importance of streaming and diffusion is strongly dependent on the pressurizing role of CRs in clouds. By pressurizing the gas, CRs reduce the cold gas density, and allow frame-locking. For instance, consider the limit where $X_{\rm CR}^{\rm hot} \sim 0.5$, as expected in CR driven winds. In this case, $P_c \rightarrow P_g^{\rm hot}$. Using equation \eqref{eqn:overdensity} for the cold gas density, we obtain: 
\begin{equation}\label{eq:vDvA3}
\frac{v_D}{v_A}-1 = 0.1  \, \frac{T_4^{1/4}(n^{\rm hot}_{-3})^2 (1+(X_{\rm CR,-0.3}^{\rm hot})^{-1})^{3} \Lambda_{-22}^{1/2}\langle\gamma\rangle}{B_{\mu G}^{3/2}P_{g,eV,-1}^{\rm hot}}.
\end{equation}
 
This has some interesting implications. Firstly, it means that the pressurizing and heating effects of CRs are crucial for CR coupling to the gas. There is a synergistic effect: if CRs can couple to the gas, they can pressurize and heat it, which makes the boundary layer between hot and cold gas thicker and less dense, which further enables frame-locking, etc. In the future, it would be interesting to explore this in time-dependent calculations where transport relative to the wave frame is incorporated by a diffusion coefficient. Secondly, it potentially implies redshift dependence in the ability of CRs to couple to multiphase gas, since from equation \eqref{eq:vDvA3}, $(v_D/v_A-1) \propto n_{h} \propto (1+z)^{3}$ at fixed virial temperature. Thus, cold gas which is coupled and pressurized by CRs at low redshift may lose coupling at high redshift, due to the higher densities which enable more efficient cooling. Another effect we have not taken into account is the fact that low density gas which is in photoionization (rather than collisional) equilibrium is over-ionized for its temperature, and has less efficient cooling $\Lambda(T)$ \citep{wiersma09}, which also enables coupling. These effects are potentially consistent with a scenario where cold gas in $z\sim 0$ halos are under-dense \citep{werk14} due to CR pressurization, but where cold gas in high redshift halos are not coupled and hence have densities consistent with thermal pressure balance \citep{lau15}.

\section{Conclusions}
\label{sec:conclusions} 
We have discussed the possible consequences of a bottleneck effect that occurs when a population of CRs undergoing streaming is incident upon a cold cloud embedded in a hot galactic halo. Even if strong wave damping in the interior of these clouds prevents coupling between the CRs and the cold gas, as CRs exit the cloud they will recouple, and at some point there will be a minimum CR drift speed. This point will serve as a bottleneck for all CRs traveling through the cloud and cause a CR pressure gradient to build up. This has a number of interesting consequences, which can be separated into large and small-scale effects.

On large scales, the bottleneck potentially induces: 
\begin{itemize} 
\item{{\bf Modulation of upstream CR profile.} The CR bottleneck flattens the upstream CR profile, which eliminates heat and momentum transfer. In particular, since upstream CRs are streaming sub-Alfv\'enically with respect to the local Alfv\'en speed in the ambient hot gas, they do not satisfy conditions for the streaming instability and are not coupled to the gas; they are only coupled downstream of the bottleneck. This highlights the global effects that small scale multi-phase structure can have on global CR distributions, with important consequences for CR-driven galactic winds.}

\item{{\bf Pushing and heating cold clouds.} The CR gradient itself pushes on the cloud material, rarefying the material inside and accelerating it to potentially high velocities for a strong CR source. Given that quasar absorption line observations frequently see cold gas outflowing at high velocities comparable to the escape velocity, in situations where entrainment by hot gas is problematic \citep{zhang15,scannapieco15}, this is a potentially important effect. We described this effect with 1D plane symmetric, time-dependent simulations using the ZEUS hydrodynamic code. It remains to be seen if such effects would hold up in 3D MHD with field-aligned CR transport.}

\end{itemize} 

On small scales, effects include: 
 
\begin{itemize}

\item{{\bf Pressurizing fronts.} CRs can pressurize cold gas. In the wave locking approximation, $P_c \propto v_A^{-\gamma_{c}}$, the relative importance of CRs increases greatly in cold gas. Since cold gas is thus dominated by non-thermal pressure, this greatly reduces the density contrast between cold and hot gas: $n^{\rm cold}/n^{\rm hot} \approx (1+(X_{\rm CR}^{\rm hot})^{-1})^{3/2}$, where $X_{\rm CR}^{\rm hot}=P_c/P_g^{\rm hot}$ is the relative CR energy density in the hot gas. Since $X_{\rm CR}^{\rm hot} \sim \mathcal{O}(1)$ in CR-driven winds, this implies that the cold and hot gas can potentially be of comparable density, consistent with recent observations \citep{werk14}. In particular, the lower density of cold gas means that the gas is more easily photoionized.} 

\item{{\bf Heating fronts.} In the absence of heat transport processes, the cold and hot gas interface is a contact discontinuity. In practice, processes like thermal conduction and turbulent mixing broaden the interface. The co-existence of gas of different temperatures at the same interface is attractive since multiple ionic species are often observed to have the same kinematic structure. CR heated fronts are a new and attractive alternative. CR heating does not have the strong temperature dependence of thermal conduction; also, since CRs pressurize fronts, reducing gas densities and cooling rates, temperature scale heights are larger at low temperatures. This increases the ratio of low to high ionic species, in better agreement with observations.}  

\end{itemize} 
Both of these effects only hold if CRs are able to couple effectively to the cold gas via a high scattering rate, which effectively requires that the parameter $l_{\rm CR}/L_{\rm CR}) (c/v_A) < 1$ (if this condition is satisfied, then the fluid approximation $\lambda_{\rm CR}/L_{\rm CR} \ll 1$ is immediately satisfied as well). Interestingly, we find that strong coupling and CR pressurization and heating have a synergistic effect -- coupling is of course required for momentum and heat transfer, but the latter two are also required for  tight coupling, by reducing densities and thickening the thermal interface.

This paper is an exploratory first effort, highlighting the many hitherto unexplored consequences of the bottleneck effect, as CRs propagate through multi-phase gas. Much more detailed study is required for a fuller understanding (before, for instance, one can furnish subgrid prescriptions for simulations of galaxy feedback). The largest uncertainty in our calculations is the degree to which CRs penetrate cold clouds. In principle, magnetic draping of field lines could insulate cold gas against CRs entering from the hot medium, rendering the bottleneck effect moot. Some degree of penetration must take place, since we see CR Coulomb heating and ionization even in dense molecular clouds \citep{padovani09,papadopoulos10}. Also, the potentially very large covering factor of cold gas (McCourt et al, 2016, in preparation) implies a high likelihood that CRs will have to stream through a cold cloud before exiting a galaxy.  Also, we have assumed that B-field strength is roughly independent of density for coronal and atomic gas, consistent with Zeeman observations in the ISM \citep{crutcher10}. If this is no longer true in halo gas, and B-field strength increases in cold gas, then magnetic mirroring could potentially exclude CRs from cold clouds \citep{chandran00}.

We also again note that all the simulations and steady state models introduced here were performed in one dimension only. The magnetic field structure is not considered, and we assumed (from analytic considerations) that frame-locking is a good approximation, when non-linear Landau damping dominates. Full 3D simulations that include the effects of different magnetic field orientations and wave damping mechanisms are necessary to verify the claims made here. 3D MHD calculations are particularly necessary to assess the importance of bulk effects, such as modification of global CR profiles and acceleration of cold clouds. The stability of clouds to being shredded by CR pressure gradients, evaporated by CR heating, and the influence that CRs have on B-field orientation can only be clarified by such simulations. As for our front calculations, while 1D steady state calculations are probably adequate if the CR scale height $L_{\rm CR}$ is small compared to the size of the cloud $R$, 3D time-dependent MHD calculations are needed once that approximation fails, and to assess if any dynamical instabilities arise in CR dominated fronts. Furthermore, incorporating photoionization and CR diffusion are essential for more robust predictions of ionic column densities in CR fronts. It would also be very interesting to study thermal instability in a CR heated medium, and in particular if there is a characteristic scale analogous to the Field length. These fascinating unexplored issues should repay detailed study.

{\bf Acknowledgements.} J.W. and S.P.O. acknowledge NASA grants NNX12AG73G and NNX15AK81G for support. J.W. and E.G.Z. acknowledge the University of Wisconsin-Madison and NSF grant AST 1616037 for support. We are also happy to acknowledge useful discussions with R.A. Benjamin. We would like to thank the referee for useful suggestions and comments.

\appendix

\section{Resolution Tests}
Figure \ref{fig:res18} shows a spatial resolution test of the results described in \S \ref{sec:regimes}. The figure plots density and velocity profiles at a time of 18 Myr after the start of the simulation. The initial density profile is that of the 25 pc edged cloud.

Three different plane-parallel grids are tested. Each grid is 40 kpc wide. The width of each cell is larger than that of the previous cell by a constant factor. Grid 1 has 400 cells, each 1.014 times wider than the previous cell. This yields a cell size of 2.2 pc at the left end, increasing to 580 pc at the far right. Grid 2 has 1600 cells, each 1.002 times wider than the previous cell. This yields a cell size of 3.5 pc at the left end, increasing to 85 pc at the right. This is the grid used in the body of this paper. Grid 3 has 3200 cells, each 1.001 times wider than the previous cell. This yields a cell size of 1.8 pc at the left end, increasing to 43 pc at the right.

\begin{figure}
\includegraphics[width=9cm, trim=1cm 5cm 0cm 0cm]{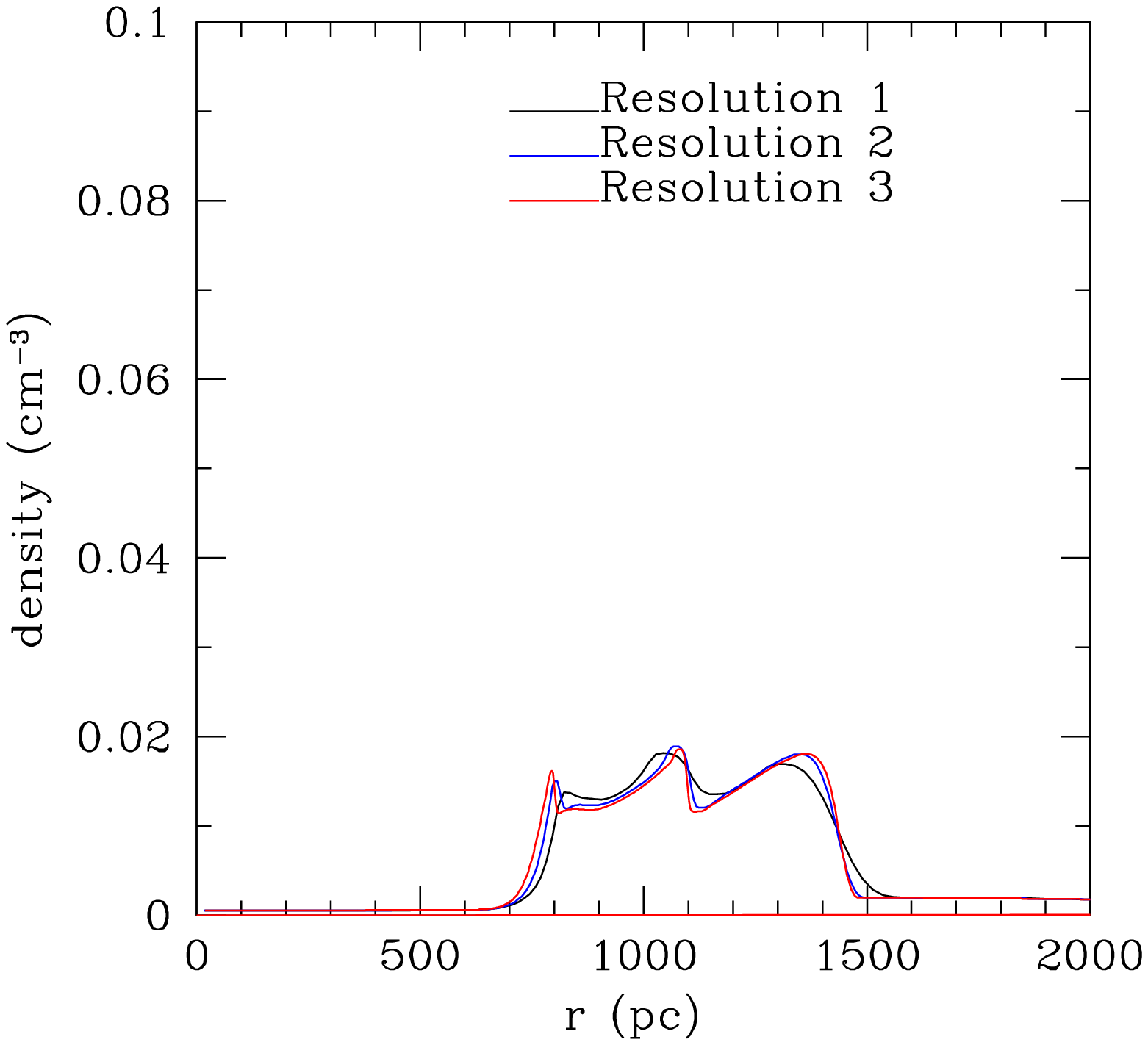}
\includegraphics[width=9cm, trim=1cm 5cm 0cm 0cm]{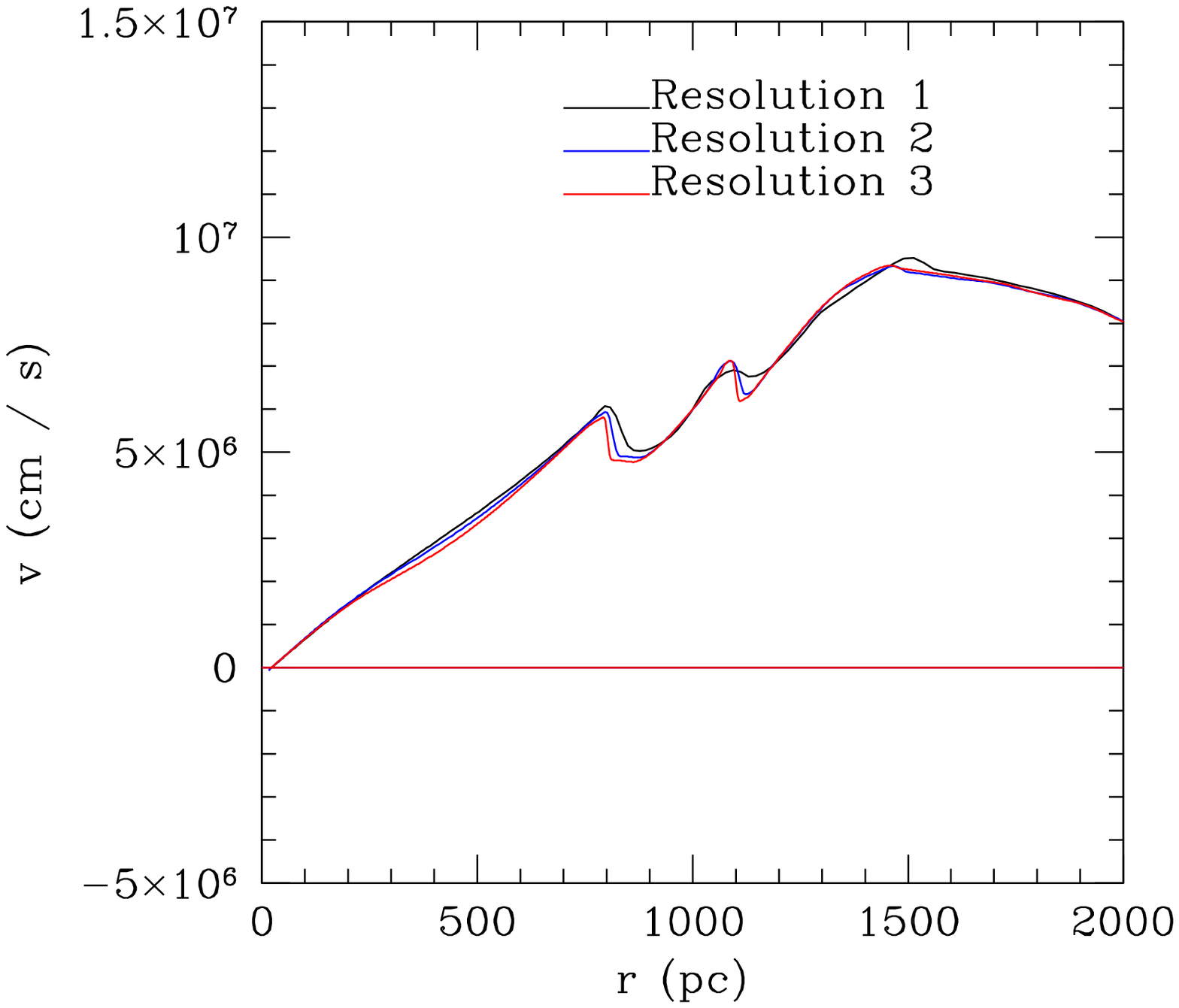}
\caption{Resolution tests for 25 pc thick edged cloud. Density and velocity profiles are shown at a time of 18 Myr. Grids are non-uniform. Reslution 1: 2 pc at left, 16 pc at 1 kpc, 60 pc at 4 kpc. Resolution 2: 3.5 pc at left, 5.5 pc at 1 kpc, 11 pc at 4 kpc. Resolution 3: 1.8 pc at left, 2.7 pc at 1 kpc, 5.7 pc at 4 kpc.}\label{fig:res18}
\end{figure}

\bibliography{master_references2}

\end{document}